\documentclass[a4paper,twocolumn,11pt,unpublished]{quantumarticle}
\pdfoutput=1
\usepackage[utf8]{inputenc}
\usepackage[english]{babel}
\usepackage[T1]{fontenc}
\usepackage{amsmath}
\usepackage{hyperref}

\usepackage{tikz}
\usepackage{lipsum}
\usepackage{subfigure}

\begin{document}

% \title{Template demonstrating the quantumarticle document class}

% \author{Lídia del Rio}
% \affiliation{Institute for Theoretical Physics, ETH Zurich, 8093 Zurich, Switzerland}
% \orcid{0000-0002-2445-2701}
% \author{Christian Gogolin}
% \email{latex@quantum-journal.org}
% \homepage{http://quantum-journal.org}
% \orcid{0000-0003-0290-4698}
% \thanks{You can use the \texttt{\textbackslash{}email}, \texttt{\textbackslash{}homepage}, and \texttt{\textbackslash{}thanks} commands to add additional information for the preceding \texttt{\textbackslash{}author}. If applicable, this can also be used to indicate that a work has previously been published in conference proceedings.}
% \affiliation{Covestro Deutschland AG, Kaiser-Wilhelm-Allee 60, 51373 Leverkusen, Germany}
% \author{Marcus Huber}
% \affiliation{Institute for Quantum Optics \& Quantum Information (IQOQI), Austrian Academy of Sciences, Boltzmanngasse 3, Vienna A-1090, Austria}
% \orcid{0000-0003-1985-4623}
% \author{Christopher Granade}
% \affiliation{Microsoft Research, Quantum Architectures and Computation Group, Redmond, WA 98052, USA}
% \author{Johannes Jakob Meyer}
% \affiliation{Dahlem Center for Complex Quantum Systems, Freie Universität Berlin, 14195 Berlin, Germany}
% \orcid{0000-0003-1533-8015}
% \author{Victor V. Albert}
% \affiliation{Institute for Quantum Information and Matter \& Walter Burke Institute for Theoretical Physics, Caltech, Pasadena, CA 91125, USA}
% \orcid{0000-0002-0335-9508}

\title{Quantum computer-based simulation of Stark many-body localization in a 1D Fermi-Hubbard model} 

% \title{Digital calculation of Stark many-body localization in a strongly correlated electronic system} 
%Alternatives? 

\author{Abdul Kalam}
\email{abdulkalam.phys@gmail.com}
\affiliation{Centre for Quantum Engineering, Research and Education, TCG CREST, Kolkata 700091, India}
\affiliation{Academy of Scientific and Innovative Research (AcSIR), Ghaziabad 201002, India}

\author{Prasenjit Deb}
% \email{devprasen@gmail.com}
\affiliation{Centre for Quantum Engineering, Research and Education, TCG CREST, Kolkata 700091, India}

\author{Akitada Sakurai}
% \email{akitada.sakurai@oist.jp}
\affiliation{Okinawa Institute of Science and Technology Graduate University, Onna-son, Okinawa 904-0495, Japan}

\author{Tapan Mishra}
% \email{mishratapan@gmail.com}
\affiliation{School of Physical Sciences, National Institute of Science Education and Research, Jatni 752050, India}

 \author{V. S. Prasannaa}
% \email{srinivasaprasannaa@gmail.com}
\affiliation{Centre for Quantum Engineering, Research and Education, TCG CREST,  Kolkata 700091, India}
\affiliation{Academy of Scientific and Innovative Research (AcSIR), Ghaziabad 201002, India}

\author{B. P. Das}
% \email{Bhanu.das@tcgcrest.org}
\affiliation{Centre for Quantum Engineering, Research and Education, TCG CREST, Kolkata 700091, India}
\affiliation{Department of Physics, Institute of Science Tokyo, 2-12-1 Ookayama, Meguro City, Tokyo 152-8550, Japan}

\maketitle

\begin{abstract}
  % In the standard, \texttt{twocolumn}, layout the abstract is typeset as a bold face first paragraph.
  % Quantum also supports a \texttt{onecolumn} layout with the abstract above the text.
  % Both can be combined with the \texttt{titlepage} option to obtain a format with dedicated title and abstract pages that are not included in the page count.
  % This format can be more suitable for long articles.
  % The \texttt{abstract} environment can appear both before and after the \texttt{\textbackslash{}maketitle} command and calling \texttt{\textbackslash{}maketitle} is optional, as long as there is an \texttt{abstract}.
  % Both \texttt{abstract} and \texttt{\textbackslash{}maketitle} however must be placed after all other \texttt{\textbackslash{}author}, \texttt{\textbackslash{}affiliation}, etc.\ commands, see also Section~\ref{sec:title-information}.
  % If you provide the ORCID number of an author by using the \texttt{\textbackslash{}orcid} command, the author name becomes a link to their page on \href{http://orcid.org/}{orcid.org}. 

Many-body localization (MBL) is a dynamical phenomenon that describes the non-ergodicity of isolated quantum many-body systems. In contrast to thermalization, this phenomenon leads to a long-lived memory of initial states of local systems and slow growth of entanglement. In this work, we study Stark MBL in a 12-qubit correlated fermionic system described by the one-dimensional Fermi–Hubbard model using Hamiltonian simulation on an IBM superconducting qubit quantum computer. To enable such a computation on current-day noisy hardware, we combine a series of compilation steps, including the use of the spin-resolved Jordan-Wigner transformation, employing SWAP networks, and integrating a tensor-network-based quantum circuit optimization routine on top of a standard circuit optimization pipeline. As a result, there is approximately an $ 88\%$ and $87 \%$ reduction in two-qubit gate count and circuit depth, respectively. Through such simulations of the real-time dynamics using Trotterized quantum circuits, we exhibit a crossover from thermalizing dynamics of the system at a weak tilt of the field to a strongly localized behavior at large tilt with short evolution times. We also benchmark our obtained results with respect to those from exact simulations. 
\end{abstract}

% In the \texttt{twocolumn} layout and without the \texttt{titlepage} option a paragraph without a previous section title may directly follow the abstract.
% In \texttt{onecolumn} format or with a dedicated \texttt{titlepage}, this should be avoided.

% Note that clicking the title performs a search for that title on \href{http://quantum-journal.org}{quantum-journal.org}.
% In this way readers can easily verify whether a work using the \texttt{quantumarticle} class was actually published in Quantum.
% If you would like to use \texttt{quantumarticle} for manuscripts not yet accepted in Quantum, or not even intended for submission to Quantum, please use the \texttt{unpublished} option to switch off all Quantum related branding and the hyperlink in the title.
% By default, this class also performs various checks to make sure the manuscript will compile well on the arXiv.
% If you do not intend to submit your manuscript to Quantum or the arXiv, you can switch off these checks with the \texttt{noarxiv} option.
% On the contrary, by giving the \texttt{accepted=YYYY-MM-DD} option, with \texttt{YYYY-MM-DD} the acceptance date, the note ``Accepted in Quantum YYYY-MM-DD, click title to verify'' can be added to the bottom of each page to clearly mark works that have been accepted in Quantum. 

\section{Introduction}\label{sec:1}

One of the primary motivations underlying the development of quantum computers is the simulation of complex quantum systems \cite{lloyd1996universal,georgescu2014quantum,preskill2018quantum}. In principle, the ability of quantum computers to simulate the dynamics of entangled many-body systems makes them suitable for exploring various quantum mechanical properties and phenomena \cite{alam2025fermionic,alam2026onset,cobos2025real}. However, to exploit the entire potential of quantum computation, building a fully programmable and fault-tolerant digital quantum processor is vital \cite{campbell2017roads,preskill2018quantum}. Though much research, theoretical and experimental, have been performed in this direction, there are numerous challenges still to be overcome \cite{battistel2023real}. Nevertheless, present-day non-fault-tolerant quantum processors, namely, \textit{noisy intermediate-scale quantum} (NISQ) devices, offer scope to capture various quantum mechanical properties of quantum many-body systems \cite{fauseweh2024quantum} and compute physical quantities in atomic and molecular systems \cite{ying2023experimental, kandala2017hardware, kandala2019error}.

One fundamental property of a generic quantum many-body system that has baffled physicists since the inception of quantum mechanics and has been an active area of interest in the field of quantum statistical mechanics is its thermalization without the aid of any heat bath. The \textit{eigenstate thermalization hypothesis} (ETH) \cite{deutsch1991quantum,srednicki1994chaos} resolves this problem by providing a framework based on the random matrix theory and quantum chaos. This hypothesis explains how the expectation value of physical observables of an isolated quantum many-body system gets reconciled with their corresponding thermodynamic averages. From ETH, it can be inferred that from an initial out-of-equilibrium state, the system evolves in such a manner that the local information encoded in its initial state gets scrambled over time, leading to thermalization, even if unaided by an external reservoir.

However, this ergodic behaviour of the system can break down in the presence of disorder, the most common one being quenched disorder, leading to the phenomenon of \textit{many-body localization} (MBL) \cite{nandkishore2015many, alet2018many, abanin2019colloquium, gopalakrishnan2020dynamics, sierant2025many, zakrzewski2026many}. Originally introduced by P. W. Anderson \cite{anderson1958absence} for a single quantum particle on a disordered crystal, this phenomenon
emerges in quantum systems with both interactions and
disorder as well. As disorder increases, a many-body system fails to thermalize even at high temperatures, and this non-ergodicity manifests in properties such as long-term memory retention, logarithmic entanglement growth over time, and area-law entanglement scaling \cite{nandkishore2015many}.
% \textcolor{red}{The transition between many-body localization and delocalization at a critical disorder strength is a dynamical phase transition that has attracted considerable attention across various fields of physics }\cite{harris2022benchmarking}.
In the last few decades, massive development in the field of computer science and technology has enabled researchers to simulate MBL of quantum many-body systems with greater precision and study various collective properties related to the dynamics of such systems \cite{nandkishore2015many}. Despite the vast amount of research on thermalization and MBL, there are still many open questions regarding the eﬀects of various factors, e.g., topology, dimensionality, long-range interactions, etc., on this phenomenon, especially in the proximity of the phase transition. Moreover, better simulations of this phenomenon would also help us to deeply understand the fundamental concepts, such as ETH, from the perspective of quantum thermodynamics. One of the most critical challenges that one may face in answering these questions theoretically and understanding these phenomena in a better way is to overcome the constraints arising due to the limitations of classical computers \cite{nandkishore2015many}. Due to this inevitable challenge in the theoretical study of many-body quantum systems, a question naturally arises -- \textit{Can the present-day quantum computers be used to study the dynamics of isolated many-body quantum systems}? From the perspective of quantum computation, the study of a many-body localized system using a quantum computer has been proposed as a benchmark for showing the utility of NISQ devices \cite{zhu2021probing}.
\paragraph*{}

While disorder has long been considered as an essential ingredient for studying MBL, recent theoretical works suggest that a quantum many-body system with a uniformly increasing linear electric field, but in the absence of disorder, can also exhibit MBL,  which is commonly known as \textit{Stark} MBL \cite{wannier1962dynamics,morong2021observation,van2019bloch, schulz2019stark, taylor2020experimental, zhang2021mobility, scherg2021observing}. In this type of localization, a linear Stark potential is used to generate a deterministic energy gradient across the lattice, suppressing particle transport and producing nonergodic dynamics at sufficiently large field strengths.  Several experimental studies exploring Stark-MBL have been carried out in various quantum many-body systems, including ultracold atoms in optical lattices \cite{wilkinson1996observation,taylor2020experimental, yao2020many, lang2022disorder} and semiconductor devices \cite{guo2021stark, wang2021stark}.
The disorder-free nature of Stark MBL is particularly advantageous for
quantum-computer simulations. In conventional MBL studies based on random
on-site disorder, observables are generally averaged over many independent
disorder realizations. For example, in  quasiperiodic models such as the
Aubry--Andr\'e model, (\(\epsilon_i=\lambda\cos(2\pi\beta n+\varphi)\)
where $\epsilon_{i}$ is the onsite potential, $\lambda$ is the disorder strength, $\beta$ is an irrational number, and $\varphi$ denotes the phase offset)
\cite{aubry1980analyticity}, which is used to create onsite disorder; observables are often averaged over several
values of $\varphi$ to reduce phase-dependent finite-size effects.
On a quantum computer, each disorder realization or phase offset requires
a separate circuit execution, together with a sufficiently large number of
measurement shots, thereby substantially increasing the experimental cost.
In contrast, the deterministic linear potential underlying Stark MBL
does not require phase averaging, which reduces the
number of circuit executions needed to characterize the dynamics.

In this article, we study Stark MBL in a correlated fermionic system, represented by one dimensional fermi Hubbard model \cite{hubbard1963electron}, using Hamiltonian simulation on a quantum computer. 
To enable such a computation on current-day noisy hardware, we integrate a tensor-network-based quantum circuit optimization routine on top of a standard circuit optimization pipeline. Localization is characterized from the evolution of various observables, and the accuracy of our analysis is verified through the agreement between the results obtained from exact diagonalization, noiseless simulation, and quantum hardware executions.
The rest of the article is arranged as follows: in Section (\ref{sec:2}), we discuss the theory and model related to our work. In Section (\ref{sec:3}), we demonstrate the results. Finally, we conclude our work in Section (\ref{sec:4}).

%%%% main flowchart with the lattice %%%%%%
% \begin{figure*}[tbh]
% \centering
% \subfigure[]{\includegraphics[width=0.40\textwidth]{All_Figures/figure1a.png}}
% \subfigure[]{\includegraphics[width=0.42\textwidth]{All_Figures/figure1b.png}}
% \caption{
% (a) Schematic representation of the tilted Fermi--Hubbard model considered in this work. The hopping amplitude $J$, on-site interaction $U$, and linear Stark potential of strength $\Delta_0$ define the disorder-free lattice model. A representative charge-density-wave (CDW) initial state and the qubit ordering used in the simulations are also shown. 
% (b) Workflow of the quantum-simulation protocol. The tilted Fermi--Hubbard model is mapped to qubits using a spin-resolved Jordan--Wigner transformation, evolved using a Trotterized SWAP-network circuit, and optimized through AQC-Tensor, Qiskit level-3, and pytket routines before hardware execution. Error suppression is performed using dynamical decoupling and Pauli twirling, followed by postselection based on particle-number and spin-sector conservation. The measured observables are then used to identify finite-time signatures of Stark-MBL.
% }

% \label{fig:figure of tilted fhm and flowchart}
% \end{figure*} 

\begin{figure*}[tbh]
    \centering
    \includegraphics[width=0.95\linewidth]{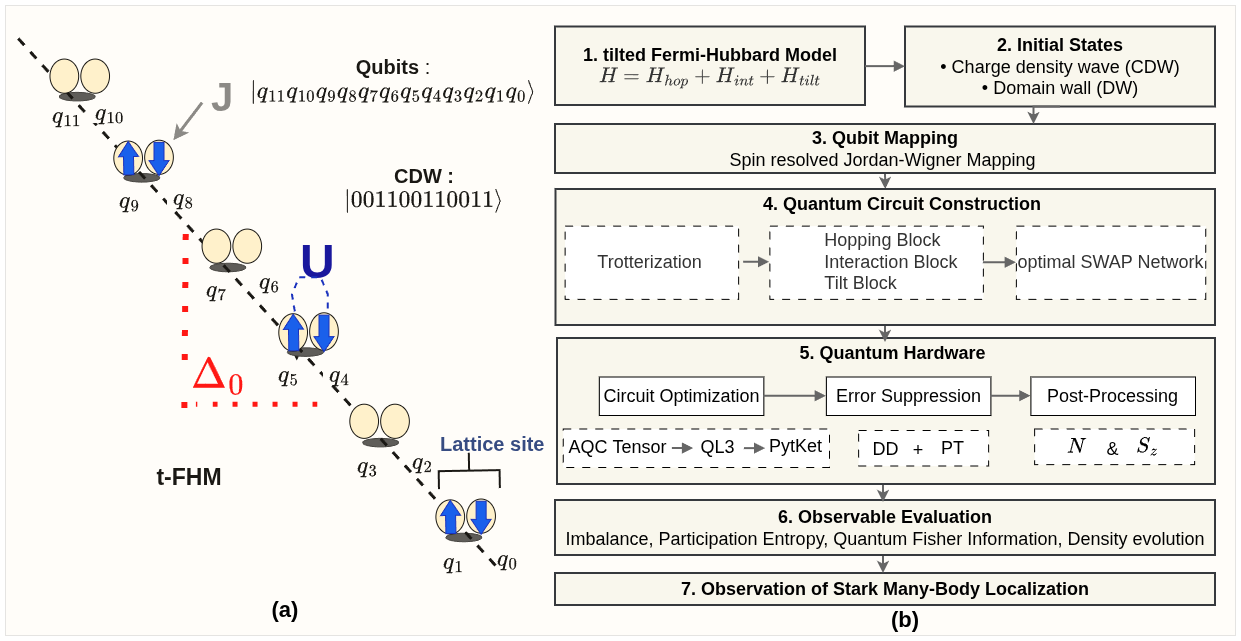}
    \caption{(a) Schematic representation of the tilted Fermi--Hubbard model considered in this work. The hopping amplitude $J$, on-site interaction $U$, and linear Stark potential of strength $\Delta_0$ define the disorder-free lattice model. A representative \textit{charge-density-wave} (CDW) initial state and the qubit ordering used in the simulations are also shown. 
(b) Workflow of the quantum-simulation protocol. The tilted Fermi--Hubbard model is mapped to qubits using a spin-resolved Jordan--Wigner transformation, evolved using a Trotterized SWAP-network circuit, and optimized through AQC-Tensor, Qiskit level-3, and pytket routines before hardware execution. Error suppression is performed using dynamical decoupling (DD) and Pauli twirling (PT), followed by postselection based on particle-number and spin-sector conservation. The measured observables are then used to identify finite-time signatures of Stark-MBL.}
    \label{fig:figure of tilted fhm and flowchart}
\end{figure*}

\section{Theory and Model}\label{sec:2}
\subsection{t-FHM}
The standard Fermi-Hubbard model \cite{hubbard1963electron,stanisic2022observing} considers electron hopping only between nearest-neighbor lattice sites and the Coulomb interaction between two electrons of opposite spin at the same site. The Hamiltonian of the underlying system is expressed as, $\hat{H}_{FH} = H_{\mathrm{hop}} + H_{\mathrm{int}}$,
\begin{equation}
\hat{H}_{FH} = -J\sum_{i,j}\sum_{\sigma}\left(\hat{a}_{i\sigma}^{\dagger}\hat{a}_{j\sigma} + \hat{a}_{j\sigma}^{\dagger}\hat{a}_{i\sigma}\right) + U\sum_{i}\hat{n}_{i\uparrow}\hat{n}_{i\downarrow}\,,
\label{eq:Hamiltonian}
\end{equation}
where $\sigma \in \{\uparrow, \downarrow\}$ is the Pauli spinor, $\hat{n}_{i\sigma} = \hat{a}_{i\sigma}^{\dagger}\hat{a}_{i\sigma}$ is the number operator for electrons with spin $\sigma$, and $\hat{a}_{i\sigma}^{\dagger}$ and $\hat{a}_{j\sigma}$ are creation and annihilation operators, respectively. 
% The term $\hat{a}_{i\sigma}^{\dagger}\hat{a}_{j\sigma}$ represents the annihilation of an electron with spin $\sigma$ at site $j$ and the creation of an electron with spin $\sigma$ at site $i$, describing hopping of electrons between $i^{th}$ and $j^{th}$ sites. The summation  $\sum_{i,j}\sum_{\sigma}$ accounts for all nearest-neighbor sites and both spin states. 
Here, $J$ is the hopping amplitude that determines the probability of electron hopping between two lattice sites and is related to the kinetic energy of electrons.
% Finally, $\hat{n}_{i\uparrow}\hat{n}_{i\downarrow}$ denotes the number of electron pairs with opposite spins on the $i^{th}$ site (taking values of 0 or 1) 
and $U$ is the Coulomb interaction potential between the two electrons of opposite spin at the same lattice site.

To investigate MBL in the absence of quenched disorder, we subject the one-dimensional Fermi-Hubbard model to a uniform external electric field, specifically, a Stark field. Our analysis considers the open boundary conditions. The resulting linear potential generates a tilt across the lattice and gives rise to the so-called \textit{tilted Fermi-Hubbard model} (t-FHM). At sufficiently strong field strengths, the interplay between the Stark potential and particle interactions suppresses transport and thermalization, leading to the phenomenon of Stark-MBL.

The Hamiltonian of the t-FHM is given by
\begin{eqnarray}
\hat{H}
&=&
\hat{H}_{{FH}}
+
\hat{H}_{{tilt}}
\nonumber  \\ 
&=&\hat{H}_{{FH}}
+
\sum_{i=0}^{L-1}
\Delta_0 i
\left(
\hat{n}_{i\uparrow}
+
\hat{n}_{i\downarrow}
\right),
\label{eq:tilted_fhm}
\end{eqnarray}
% &=&
% -J \sum_{\langle i,j \rangle,\sigma}
% \left(
% \hat{a}_{i\sigma}^{\dagger}\hat{a}_{j\sigma}
% +
% \hat{a}_{j\sigma}^{\dagger}\hat{a}_{i\sigma}
% \right)
% \nonumber\\
% &&+
% U\sum_i
% \hat{n}_{i\uparrow}\hat{n}_{i\downarrow}
% \nonumber\\
% &&+
% \sum_{i=0}^{L-1}
% \Delta_0 i
% \left(
% \hat{n}_{i\uparrow}
% +
% \hat{n}_{i\downarrow}
% \right),
% \label{eq:tilted_fhm}
% \end{eqnarray}
where the final term corresponds to the Stark potential and $\Delta_0$ characterizes the strength of the applied electric field. This term generates a linear energy gradient across the lattice, effectively producing the tilt responsible for Stark localization.
A pictorial representation of the t-FHM is shown in Figure~\ref{fig:figure of tilted fhm and flowchart}(a). 
The Hamiltonian in Eq.({\ref{eq:tilted_fhm}) commutes with the particle number operator
$\hat{N}$ and the $z$-component of the total spin operator
$\hat{S}_{z}$ in the system \cite{arovas2022hubbard, grosse1989symmetry}, i.e., $[H,\hat{N}_{\mathrm{tot}}]=[H,\hat{S}^{z}_{\mathrm{tot}}]=0\,$,
where $\hat{N} = \sum_{i,\sigma} \hat{a}_{i\sigma}^{\dagger}\hat{a}_{i\sigma}$ and $\hat{S}_{z} = \frac{1}{2} \sum_{i} \left(\hat{n}_{i\uparrow} - \hat{n}_{i\downarrow}\right)$.
To simulate the model on a quantum computer, the fermionic Hamiltonian in Eq.~(\ref{eq:tilted_fhm}) is qubitized using a spin-resolved Jordan-Wigner transformation as described in Ref. \cite{chowdhury2026quantum, jordan1928paulische}. The transformed Hamiltonian thus looks like, 
\begin{align}
H
&=
-\frac{J}{2}
\sum_{j=0}^{L-2}
\Big(
X_{2j}X_{2j+2}
+
Y_{2j}Y_{2j+2}
\nonumber\\
&\qquad\qquad
+
X_{2j+1}X_{2j+3}
+
Y_{2j+1}Y_{2j+3}
\Big)
\nonumber\\
&\quad
+
\frac{U}{4}
\sum_{j=0}^{L-1}
\Big(
I_{2j}I_{2j+1}
+
Z_{2j}Z_{2j+1}
\nonumber\\
&\qquad\qquad
-
Z_{2j}
-
Z_{2j+1}
\Big)
\nonumber\\
&\quad
+
\sum_{j=0}^{L-1}
\frac{\Delta_0 j}{2}
\Big(
2I
-
Z_{2j}
-
Z_{2j+1}
\Big).
\label{eq:qubit_hamiltonian}
\end{align}
During the mapping of the Hamiltonian, each lattice site is represented by two qubits corresponding to the spin-up and spin-down degrees of freedom,
$(j,\uparrow)\longrightarrow q_{2j}$ and $(j,\downarrow)\longrightarrow q_{2j+1}$ such that a lattice containing $L$ sites is encoded using $2L$ qubits. The local basis states
$\{|00\rangle,|10\rangle,|01\rangle,|11\rangle\}$
represent an empty site, a spin-up fermion, a spin-down fermion, and a doubly occupied site (spin-up and spin-down), respectively.
Unlike the conventional interleaved Jordan-Wigner mapping, where all spin sites are placed on a single parity chain, in our study, we use a spin-resolved ordering in which the spin-up and spin-down sectors are mapped onto separate parity chains. The spin-up fermions are assigned to even-indexed qubits and the spin-down fermions to odd-indexed qubits. The mapping scheme adopted by us preserves the fermionic anti-commutation relations while reducing the length of the Pauli strings associated with hopping processes.

The spin-resolved ordering of the fermions used in the mapping process provides a useful advantage in making the hopping operators shorter. In an interleaved Jordan--Wigner chain, same-spin hopping can generate strings such as $XZX+YZY$ because of the intermediate parity operator. In the spin-resolved mapping, the corresponding nearest-neighbor hopping within each spin sector is reduced to shorter $XIX+YIY$-type operators, with the identity on the opposite-spin qubit left implicit. The on-site interaction and Stark-potential terms remain diagonal and are represented only by local $Z$ and $ZZ$ operators. This gives a compact qubit representation of the t-FHM that is better suited for shallow-circuit implementation on quantum hardware.

\begin{figure*}
    \centering
    \includegraphics[width=1\linewidth]{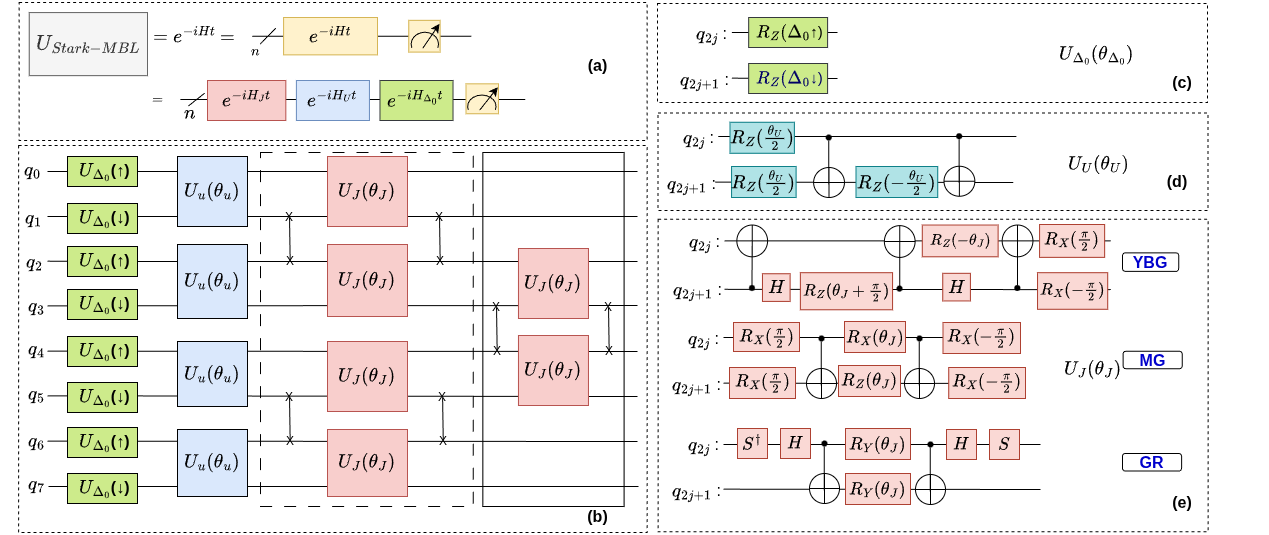}
    \caption{
First-order Trotter implementation of the t-FHM. (a) The time-evolution operator is decomposed into hopping, interaction, and Stark-potential contributions. (b) Single Trotter-step circuit employing a constant-depth SWAP-network architecture, where the Stark-potential, interaction, and hopping blocks are applied sequentially. The SWAP network enables efficient implementation of the next-nearest-neighbor hopping interactions arising from the spin-resolved Jordan--Wigner mapping using only nearest-neighbor connectivity. (c) Three alternative realizations of the hopping unitary $U_J(\theta_J)$ considered in this work: Yang--Baxter gate (YBG), Matchgate (MG), and Fermionic Givens Rotation (GR). (d) Circuit decomposition of the on-site interaction unitary $U_U(\theta_U)$. (e) Circuit decomposition of the Stark-potential unitary $U_{\Delta_0}(\theta_{\Delta_0})$.
}
    \label{fig:single_trotter_step}
\end{figure*}

\subsection{Initial States}

For the t-FHM, we consider two experimentally relevant product states \cite{schreiber2015observation} as initial states: a \textit{charge-density-wave} (CDW) state and a \textit{domain-wall} (DW) state. The CDW state is defined as
\begin{equation}
|\psi_{\mathrm{CDW}}(0)\rangle
=
|\uparrow\downarrow\,0\,0\,\uparrow\downarrow\,0\,0\,\cdots\rangle,
\end{equation}
where doubly occupied sites are separated by empty sites. 
% This state is useful for probing the persistence of charge order through the imbalance.
The DW state is prepared by occupying one half of the lattice and leaving the other half empty,
\begin{equation}
|\psi_{\mathrm{DW}}(0)\rangle
=
|\uparrow\downarrow\,\uparrow\downarrow\,\uparrow\downarrow\,\cdots\,0\,0\,0\,\cdots\rangle .
\end{equation}
It provides a direct probe of particle transport through the density profile and the particle number transferred across the center of the chain. Since both states are simple product states, they provide clear initial conditions for distinguishing between ergodic and Stark-localized dynamics.

\subsection{Trotterized Time Evolution}
The nonequilibrium dynamics of the t-FHM are simulated using the Trotter--Suzuki decomposition \cite{trotter1959product, suzuki1976generalized}, which approximates the exponential of a sum of noncommuting Hamiltonian terms by a product of exponentials acting sequentially. The exact time-evolution operator is

\begin{equation}
U(t)=e^{-iHt},
\end{equation}
which, in the first-order Trotter approximation, can be expressed as
\begin{equation}
U(t) =\lim_{n\rightarrow\infty}
\left(
\prod_{\alpha}
e^{-iH_{\alpha}t/n}
\right)^n,
\label{eq:trotter}
\end{equation}
where $\Delta t=t/n$ is the Trotter time step.
The qubit Hamiltonian is naturally decomposed into hopping, interaction, and Stark-potential contributions, $H
= H_{\mathrm{hop}}
+
H_{\mathrm{int}}
+
H_{\mathrm{tilt}},$
for a single Trotter step
\begin{equation}
U_{\mathrm{Trotter}}(\Delta t)
=e^{-iH_{\mathrm{hop}}\Delta t}
e^{-iH_{\mathrm{int}}\Delta t}
e^{-iH_{\mathrm{tilt}}\Delta t}
+ \mathcal{O}(\Delta t^2).
\end{equation}
Repeated application of this operator generates the approximate dynamics up to the desired evolution time. For an $p$th-order product formula with $n$ time steps, the Trotter error scales as
$\varepsilon
=
\mathcal{O}
\left(
\frac{t^{p+1}}{n^p}
\right)
\label{eq:trotter_error}$.
Increasing the number of Trotter steps therefore improves the accuracy of the simulation by reducing the trotter error. However, this improvement comes at the cost of a deeper quantum circuit since the number of gate operations grows proportionally with the number of time slices.

Following the strategy introduced in Refs.~\cite{chowdhury2024enhancing,chowdhury2026quantum}, we decompose the evolution operator into separate hopping, interaction, and Stark-potential blocks and implement the hopping terms through a SWAP-network. This approach enables efficient simulation of the next-nearest-neighbor interactions appearing in the spin-resolved Jordan--Wigner Hamiltonian while maintaining compatibility with the nearest-neighbor connectivity constraints of current superconducting quantum processors.

\subsection{Circuit Construction}
Figure~\ref{fig:single_trotter_step} illustrates the structure of a single first-order Trotter step used in our real-time simulation. The circuit is organized into three physically distinct blocks: the Stark-potential block, the on-site interaction block, and the hopping block. The Stark block implements the linear potential generated by the external electric field, the interaction block accounts for the local Coulomb repulsion between spin-up and spin-down fermions occupying the same lattice site, and the hopping block describes particle transport between neighboring lattice sites. These blocks are applied sequentially and repeated for the required number of Trotter steps to simulate the time evolution of the t-FHM.

To reduce the circuit gate counts, we use the SWAP-network construction. The network consists of alternating layers of SWAP gates that rearrange the qubit ordering such that the fermionic modes participating in a given hopping operation are brought next to each other. In the first SWAP layer, gates are applied between qubits $(2j+1)$ and $(2j+2)$, with $j
=0,1,\ldots,
\frac{N-2}{2},$
where $N=2L$ is the total number of qubits for a lattice of $L$ sites. The subsequent layers alternate the pairing pattern, allowing all required hopping terms to be implemented while preserving a parallel circuit structure. As the lattice size increases, the number of gates within each layer grows, but mutually disjoint gates can be executed simultaneously. Therefore, the depth contribution of a single SWAP-network is independent of system size, whereas the total circuit depth is governed primarily by the number of Trotter steps.

The circuit implementations of the Stark potential, on-site interaction, and hopping terms are shown in Figure~\ref{fig:single_trotter_step}. Since both the Stark and interaction Hamiltonians remain diagonal after the spin-resolved Jordan--Wigner transformation, they are implemented using local $R_Z$ rotations and nearest-neighbor $ZZ$ evolutions. As a result, these blocks contribute only modestly to the overall circuit depth.
The hopping term, generated by the kinetic Hamiltonian $H_{\mathrm{hop}}$, gives rise to an $XX+YY$ interaction between qubits belonging to the same spin sector. Because these qubits are separated in the spin-resolved ordering, a SWAP-network construction is employed to bring the relevant qubit pairs together before applying the hopping evolution. We consider three implementations of the hopping unitary $U_{\mathrm{hop}}(\theta_J)$: the \textit{Yang--Baxter gate} (YBG)~\cite{zhang2024optimal}, \textit{Matchgate} (MG)~\cite{bassman2022constant}, and \textit{Fermionic Givens Rotation} (GR)~\cite{anselmetti2021local} decompositions. Although these constructions differ in their gate structure, they realize the same particle-number-conserving hopping dynamics. 
Benchmark calculations in the simulation showed nearly identical fidelities and physical observables for all three decompositions. Consequently, because of the limited hardware calculation time, the MG unitary implementation is used for the hardware calculations and all the comparisons.

\subsection{Circuit Optimization}

Circuit optimization is performed using a three-step transpilation workflow. The circuits are first optimized using the tensor-network-based circuit compression method implemented using AQC-Tensor~\cite{robertson2025approximate}. For each evolution time $t$, the Trotterized circuit defines a target state $|\psi_{\mathrm{target}}(t)\rangle
=
U_{\mathrm{Trotter}}(t)
|\psi_0\rangle .$
AQC-Tensor constructs a parameterized compressed circuit,

\begin{equation}
|\psi_{\mathrm{AQC}}(\boldsymbol{\theta},t)\rangle
=U_{\mathrm{AQC}}(\boldsymbol{\theta},t)
|\psi_0\rangle ,
\end{equation}

and optimizes its parameters using the L-BFGS optimizer \cite{liu1989limited} to maximize the state fidelity

\begin{equation}
F(\boldsymbol{\theta})
=
\left|
\left\langle
\psi_{\mathrm{target}}(t)
\middle|
\psi_{\mathrm{AQC}}(\boldsymbol{\theta},t)
\right\rangle
\right|^2 .
\end{equation}

The overlap is evaluated using tensor-network contractions within the AQC-Tensor framework. A compressed circuit is accepted only when the optimized fidelity exceeds $0.9999$, and the two-qubit gate count is reduced relative to the original Trotter circuit. Since the optimization is performed at the level of the final quantum state, the underlying physical dynamics remain unchanged.

The compressed circuits are subsequently passed through the Qiskit optimization level-3 (L3)~\cite{javadi2024quantum}, followed by pytket optimization passes~\cite{sivarajah2020t}, and finally a second Qiskit L3 optimization stage. This optimization pipeline further reduces circuit depth, routing overhead, and two-qubit gate counts while preserving the target-state fidelity. The final optimized circuits are then used for execution on quantum hardware.

\subsection{Localization-detecting Measures}

To characterize the nonequilibrium dynamics of the t-FHM and identify signatures of Stark-MBL, we evaluate several observables during real-time evolution. Stark-MBL is reflected by the retention of memory of the initial state, suppression of particle transport, slow growth of entanglement, and restricted spreading of the wave function in Hilbert space.
% The time-evolved state is given by $|\psi(t)\rangle =
% e^{-i\hat{H}t}
% |\psi(0)\rangle ,$
% where $|\psi(0)\rangle$ is either the CDW or DW state. All observables discussed below are evaluated using the evolved state $|\psi(t)\rangle$.

\paragraph{Charge Imbalance :}
The memory retained from the initial CDW pattern is quantified through charge imbalance which is one of the most widely used probes of localization. It is defined as

\begin{equation}
\mathcal{I}_{c}(t)
=
\frac{N_e(t)-N_o(t)}
{N_e(t)+N_o(t)},
\end{equation}

% where $N_e(t)=\sum_{i\in \mathrm{even}}
% \langle n_i(t)\rangle,
% \qquad
% N_o(t)=\sum_{i\in \mathrm{odd}}
% \langle n_i(t)\rangle$,
% and
% $n_i(t)
% =n_{i\uparrow}(t)
% +
% n_{i\downarrow}(t)$. 
where
\(N_{\mathrm{e}}(t)
= \sum_{j\in \mathrm{even}}
\langle \psi(t)|\hat n_j|\psi(t)\rangle,
N_{\mathrm{o}}(t)
=\sum_{j\in \mathrm{odd}}
\langle \psi(t)|\hat n_j|\psi(t)\rangle .\)
Here, \(N_{\mathrm{e}}(t)\) and \(N_{\mathrm{o}}(t)\) denote the total particle number on the even and odd lattice sites, respectively. The local density operator at site \(j\) is \(\hat n_j
=
\hat n_{j\uparrow}
+
\hat n_{j\downarrow},\)
where \(\hat n_{j\sigma}=\hat a^{\dagger}_{j\sigma}\hat a_{j\sigma}\) counts particles with spin \(\sigma\) at site \(j\). 
In the thermalizing regime, particles redistribute throughout the lattice and the imbalance decays toward zero. In contrast, within the Stark-localized regime, the system retains memory of the initial CDW configuration, leading to a finite long-time imbalance.

\paragraph{Density Evolution :}

To directly visualize particle transport, we calculate the local density evolution

\begin{equation}
    \hat{n}_i(t) = \langle \psi(t) | \hat{n}_i |\psi(t)\rangle,
    \label{eq:density_op}
\end{equation}
where $\hat{n}_i$ denotes the density operator at lattice site $i$, $\hat n_i=\hat n_{i\uparrow}+\hat n_{i\downarrow}$ is the total density operator at lattice site $i$. In the ergodic regime, particles spread across the entire lattice, whereas in the Stark-MBL regime, the density remains localized near its initial distribution, indicating strongly suppressed transport.

\begin{table*}[htbp]
\centering
% \normalsize
\setlength{\tabcolsep}{2.5pt}
\renewcommand{\arraystretch}{1.08}
\caption{Circuit compression results for the CDW and DW initial states. The initial (unoptimized) Trotter 2q-gate (Here 2q means two qubit gates) count and depth are identical for both initial states and are shown only once. Percentage reductions are computed relative to the unoptimized Trotter circuit.}
\label{tab:circuit_gate_counts_depth}

\begin{tabular*}{\textwidth}{@{\extracolsep{\fill}} c c r r r r r r r r r r @{}}
\hline
 & & & \multicolumn{2}{c}{CDW: 2q Gates} 
     & \multicolumn{2}{c}{DW: 2q Gates} 
     & & \multicolumn{2}{c}{CDW: Depth} 
     & \multicolumn{2}{c}{DW: Depth} \\
\cline{4-5}
\cline{6-7}
\cline{9-10}
\cline{11-12}
$\Delta_{0}$ & $t$ 
& Unoptimized  
& Optimized & Red. 
& Optimized & Red. 
& Unoptimized 
& Optimized & Red. 
& Optimized & Red. \\
 & & 
& & (\%) 
& & (\%) 
& 
& & (\%) 
& & (\%) \\
\hline
1  & 0.00 & 0    & 0   & 0.00  & 0   & 0.00  & 1    & 1   & 0.00  & 1   & 0.00  \\
1  & 0.25 & 253  & 193 & 23.72 & 174 & 31.22 & 213  & 187 & 12.21 & 152 & 28.63 \\
1  & 0.50 & 505  & 229 & 54.65 & 212 & 58.01 & 433  & 219 & 49.42 & 215 & 50.34 \\
1  & 0.75 & 757  & 235 & 68.96 & 225 & 70.27 & 653  & 219 & 66.46 & 219 & 66.46 \\
1  & 1.00 & 1009 & 230 & 77.21 & 231 & 77.10 & 873  & 218 & 75.03 & 218 & 75.03 \\
1  & 1.25 & 1261 & 234 & 81.44 &  231 & 81.68 & 1093 & 219 & 79.96 & 219 & 79.96 \\
\hline
3  & 0.00 & 0    & 0   & 0.00  & 0   & 0.00  & 1    & 1   & 0.00  & 1   & 0.00  \\
3  & 0.25 & 253  & 197 & 22.13 & 175 & 30.83 & 213  & 204 & 4.23  & 149 & 30.04 \\
3  & 0.50 & 505  & 221 & 56.24 & 207 & 59.01 & 433  & 218 & 49.65 & 215 & 50.34 \\
3  & 0.75 & 757  & 232 & 69.35 & 231 & 69.48 & 653  & 219 & 66.46 & 219 & 66.46 \\
3  & 1.00 & 1009 & 234 & 76.81 & 232 & 77.01 & 873  & 218 & 75.03 & 219 & 74.91  \\
3  & 1.25 & 1261 & 235 & 81.36 & 235 & 81.36 & 1093 & 219 & 79.96 & 219 & 79.96 \\
\hline
15 & 0.00 & 0    & 0   & 0.00  & 0   & 0.00  & 1    & 1   & 0.00  & 1   & 0.00  \\
15 & 0.25 & 253  & 142 & 43.87 & 115 & 54.55 & 213  & 146 & 31.46 & 104 & 51.17 \\
15 & 0.50 & 505  & 150 & 70.30 & 138 & 72.67 & 433  & 146 & 66.28 & 138 & 68.13 \\
15 & 0.75 & 757  & 156 & 79.39 & 147 & 80.58 & 653  & 147 & 77.49 & 147 & 77.49 \\
15 & 1.00 & 1009 & 158 & 84.34 & 136 & 86.52 & 873  & 147 & 83.16 & 143 & 83.62 \\
15 & 1.25 & 1261 & 158 & 87.47 & 153 & 87.87 & 1093 & 147 & 86.55 & 146 & 86.64 \\
\hline
\end{tabular*}
\end{table*}

\begin{figure*}[]
\centering
\subfigure[]{\includegraphics[width=0.49\textwidth]{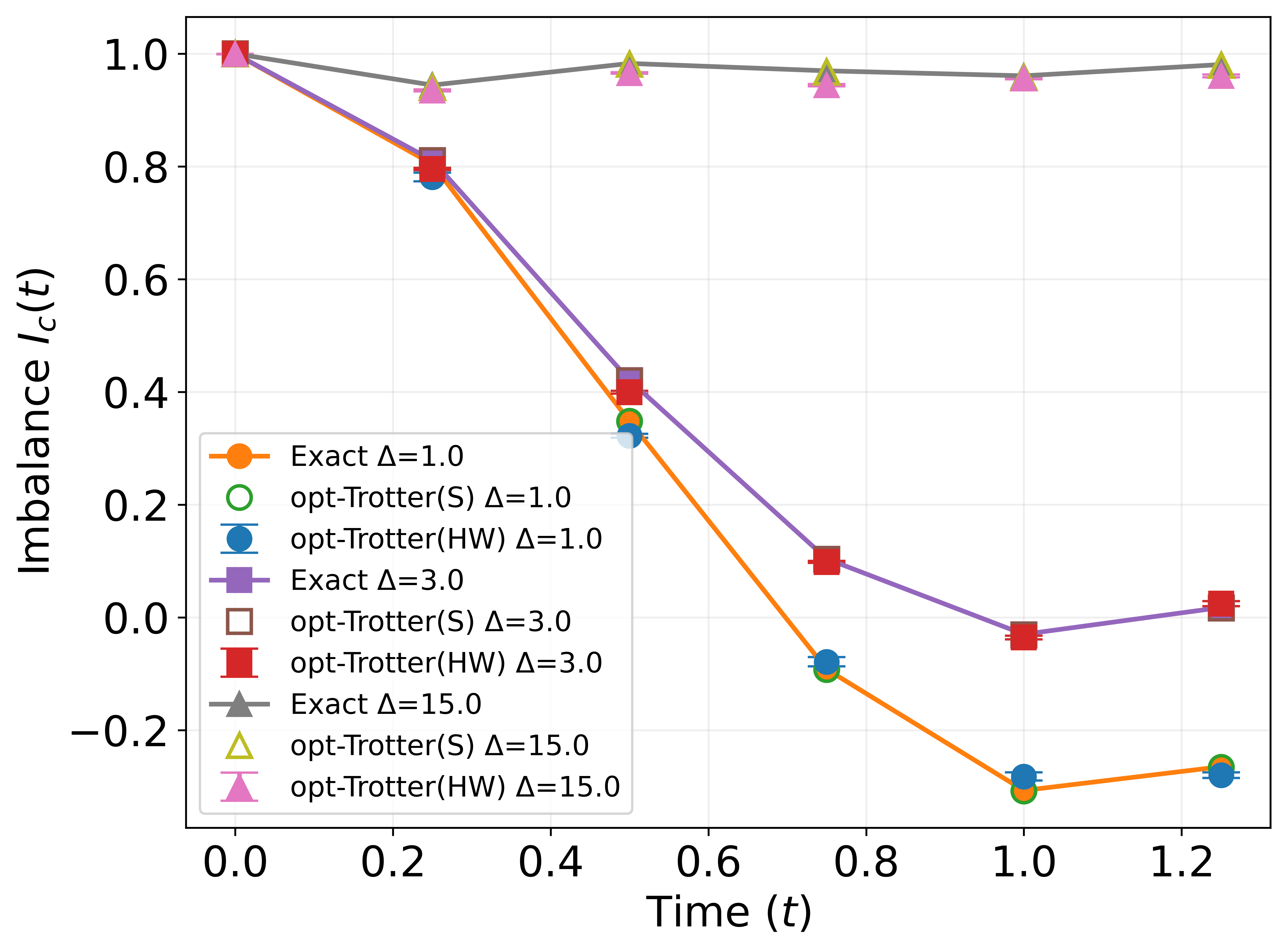}}
\subfigure[]{\includegraphics[width=0.49\textwidth]{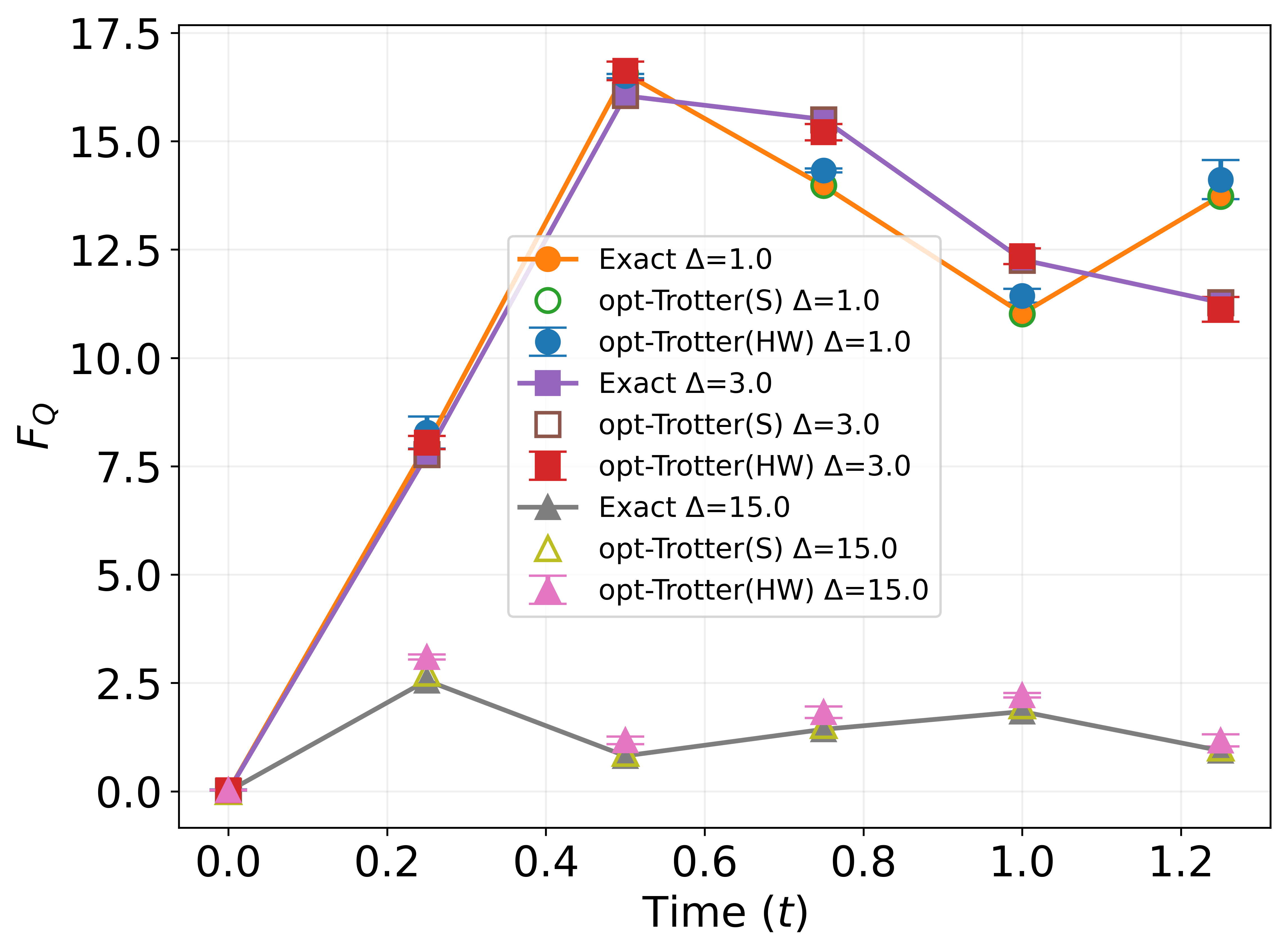}}
% \subfigure[]{\includegraphics[width=0.329\textwidth]{All_Figures/s2_entropy_june26.png}}
\caption{
Comparison of exact diagonalization (Exact), optimized Trotter simulation [opt-Trotter(S)], and optimized Trotter hardware [opt-Trotter(HW)] results for a $L=6$ t-FHM chain initialized in the CDW state. The optimized Trotter circuits are obtained after the circuit-compression and transpilation workflow described in Sec.~\ref{sec:3}. Results are shown for three representative Stark-field strengths: $\Delta_0=1$ (weak tilt), $\Delta_0=3$ (intermediate tilt), and $\Delta_0=15$ (strong tilt). (a) Charge imbalance $\mathcal{I}_c(t)$, and (b) Quantum Fisher Information $F_Q(t)$. For weak and intermediate tilts, the imbalance rapidly decays while $F_Q$ and $S_2^{\mathrm{PE}}$ increase, indicating delocalized dynamics. In contrast, the strong-tilt regime preserves a large charge imbalance and exhibits strongly suppressed growth of both $F_Q$ and $S_2^{\mathrm{PE}}$, consistent with Stark-MBL. The optimized hardware results closely reproduce the optimized noiseless Trotter simulation and remain in good agreement with the exact dynamics.}
\label{fig:imbalance_qfi_s2}
\end{figure*} 

\paragraph{von Neumann Entanglement Entropy :}

Entanglement growth provides important insight into the dynamical properties of interacting many-body systems. For a subsystem $A$, the reduced density matrix is obtained by tracing out the remaining degrees of freedom,

\begin{equation}
\rho_{A}
=\mathrm{Tr}_{A}
\left(
|\psi(t)\rangle\langle\psi(t)|
\right).
\end{equation}
The corresponding von Neumann entropy is
\begin{equation}
S_A(t)
=-\mathrm{Tr}
\left[
\rho_A
\log_2 \rho_A
\right].
\end{equation}

In this work, we evaluate the entanglement entropy for both half-chain bipartitions and individual subsystem partitions. The growth of entanglement entropy serves as a sensitive probe of localization. Thermalizing systems exhibit rapid entanglement growth and eventually approach a volume-law scaling, whereas localized systems show strongly suppressed growth. In Stark-MBL, the entanglement entropy typically grows much more slowly than in the ergodic phase, reflecting constrained information spreading.

\paragraph{Participation Entropy:}

While the von Neumann entropy characterizes entanglement present in the system, the participation entropy quantifies the spreading of the wavefunction in Hilbert space. The $q$th-order participation entropy is defined as

\begin{equation}
S_q^{\mathrm{PE}}(t)
=\frac{1}{1-q}
\log
\left(
\sum_i p_i^q(t)
\right),
\end{equation}

where $p_i(t)
=|\langle i|\psi(t)\rangle|^2$
denotes the probability associated with basis state $|i\rangle$.
The second-order participation entropy, $S_2^{\mathrm{PE}}(t)
=-\log
\left(
\sum_i p_i^2(t)
\right),$
is employed throughout this work. A small value of $S_2^{\mathrm{PE}}$ indicates that only a few many-body basis states participate in the dynamics, corresponding to localization in Hilbert space. Larger values imply extensive exploration of the Hilbert space and are characteristic of ergodic behavior.

\paragraph{Particle Number in Half of the Chain :}

For the DW initial state, transport can be directly monitored through the number of particles transferred across the center of the lattice. We therefore compute

\begin{equation}
N_{\mathrm{half}}(t)
=\sum_{j=1}^{L/2}
\langle
\psi(t)
|
\hat n_j
|
\psi(t)
\rangle .
\label{eq:Nhalf}
\end{equation}

Initially, particles occupy only one side of the system. In the thermal phase, particle transport causes $N_{\mathrm{half}}(t)$ to increase substantially with time. Conversely, in the Stark-localized regime, transport is strongly suppressed and $N_{\mathrm{half}}(t)$ remains close to its initial value.

\paragraph{Quantum Fisher Information :}

To quantify multipartite quantum correlations generated during the dynamics, we evaluate the \textit{Quantum Fisher Information} (QFI), which also serves as an experimentally accessible entanglement witness \cite{hyllus2012fisher,toth2012multipartite} and is closely related to the von Neumann entropy. For the staggered operator

\begin{equation}
\hat O
=\frac{1}{2}
\sum_j
c_j
\hat{\sigma}_j^z,
\end{equation}
with $c_j=+1$ on even sites and $c_j=-1$ on odd sites, the QFI is given by
\begin{equation}
F_Q(t)
=\sum_{jk}
c_j c_k
\langle
\hat{\sigma}_j^z
\hat{\sigma}_k^z
\rangle
-
\left(
\sum_j
c_j
\langle
\hat{\sigma}_j^z
\rangle
\right)^2.
\label{eq:QFI}
\end{equation}

For a system of $\mathcal{N}$ qubits, the quantum Fisher information satisfies the bound
$F_Q\le \mathcal{\mathcal{N}}^2$ for arbitrary entangled states \cite{smith2019simulating}. Moreover, the criterion
$F_Q/\mathcal{\mathcal{N}}\ge m$
certifies the presence of at least $(m+1)$-partite entanglement~\cite{smith2019simulating, morong2021observation}. In particular, the condition $F_Q/\mathcal{N}>1$ guarantees multipartite entanglement. Beyond serving as an experimentally accessible entanglement witness, the QFI quantifies the build-up of many-body correlations and provides information complementary to the von Neumann entanglement entropy. Consequently, its time evolution offers an additional probe of the crossover between ergodic and Stark-localized dynamics. The derivation of Eq.~(\ref{eq:QFI}) is provided in the Supplemental Material, appendix A.
% \textcolor{red}{what qfi low and high tells about the MBL}

%%%%%%%%%%%%%%%%%%%%%%%%%%%%%%%%%%%%%%%%%%%%%%%%%
\section{Results}
\label{sec:3}

\subsection{Hardware execution details}
In this section, we investigate the nonequilibrium dynamics of the t-FHM and characterize the emergence of Stark-MBL through a combination of classical simulations and quantum-hardware experiments. 
Classical simulations are performed using exact diagonalization in the QuSpin package \cite{weinberg2017quspin}.
The dynamics are initialized from CDW and DW product states and evolved under the Trotterized t-FHM described in Sec.~II.
We consider three representative tilt strengths corresponding to weak, intermediate, and strong Stark fields, namely $\Delta_0=1$, $\Delta_0=3$, and $\Delta_0=15$. These values enable us to investigate the crossover from delocalized dynamics to the strongly localized Stark-MBL regime. We fix the hopping amplitude and on-site interaction strength to $J=1$ and $U=1$, and vary only the Stark-field strength $\Delta_0$. This choice places the system in an interacting regime where hopping and local correlations compete on comparable energy scales, allowing the effect of the linear potential to be isolated clearly.
A full scan over both $U$ and $\Delta_0$ would provide a more complete phase diagram; however, such a study requires a substantially larger number of quantum circuits and is beyond the present hardware budget. Our aim here is therefore to demonstrate how increasing the Stark field suppresses transport and produces localization signatures for a fixed interacting Fermi--Hubbard system.
We note that larger interactions can further modify the localization dynamics, but the present choice of $U=J$ already captures the essential competition between hopping, interaction, and Stark tilt that is needed to observe disorder-free localization behavior.

The real-time dynamics are simulated for evolution times up to $t=1.25$ in the quantum-hardware experiments using a first-order Trotter decomposition. For each tilt strength, six time points are considered, corresponding to $t=0,0.25,0.50,0.75,1.00$, and $1.25$, and four Trotter steps per sampling interval are added. 

The quantum-hardware experiments are performed on IBM Quantum's \textit{IBM\_Marrakesh} processor based on the Heron r2 architecture. The device consists of 156 fixed-frequency transmon qubits arranged in a heavy-hex lattice with tunable couplers and supports the native gate set \{$\mathrm{id}$, $rz$, $rx$, $x$, $sx$, $cz$, $rzz$\}. At the time of execution, the median energy relaxation time ($T_1$) was $195.74~\mu\mathrm{s}$, the median dephasing time ($T_2$) was $106.81~\mu\mathrm{s}$, the median single-qubit $sx$ error rate was $2.834\times10^{-4}$, the median two-qubit $cz$ error rate was $2.521\times10^{-3}$, and the median readout error was $8.972\times10^{-3}$.
Each circuit was executed with 20,000 measurement shots. To reduce statistical fluctuations, every hardware experiment was repeated three times, and the final reported value was obtained by averaging over the independent runs. For each tilt strength, six circuits corresponding to different evolution times were executed, resulting in a total of 54 hardware circuit executions across all tilt values.

To improve the quality of the hardware results, we employ \textit{dynamical decoupling} (DD) \cite{viola1999dynamical} and \textit{Pauli twirling} (PT) \cite{bennett1996purification} as error-suppression techniques. DD reduces decoherence and unwanted spectator-qubit interactions by applying refocusing pulses during idle periods of the circuit, thereby extending the effective coherence time of the computation~\cite{viola1999dynamical,ezzell2022dynamical}. In this work, DD is implemented using the sequence
$\left(
\frac{t}{4},
X,
\frac{t}{2},
X,
\frac{t}{4}
\right),$
where $X$ denotes an $\texttt{XGate}$ and $t$ is the idle duration excluding pulse execution times.
Pauli twirling is used to convert coherent-gate errors into an effectively stochastic noise channel by randomly inserting Pauli operators before and after Clifford operations, while preserving the logical action of the circuit~\cite{bennett1996purification,wallman2016noise,cai2019constructing}. This transformation suppresses coherent error accumulation and improves the robustness of noisy quantum simulations. 
% In our implementation, PT is applied to all $\texttt{CZ}$ gates. For each hardware run, a twirling sequence is randomly selected from the set of Pauli combinations that leave the $\texttt{CZ}$ operation unchanged up to a global phase~\cite{Kim-error-mitigation}.

\begin{figure}
    \centering
    \includegraphics[width=0.95\linewidth]{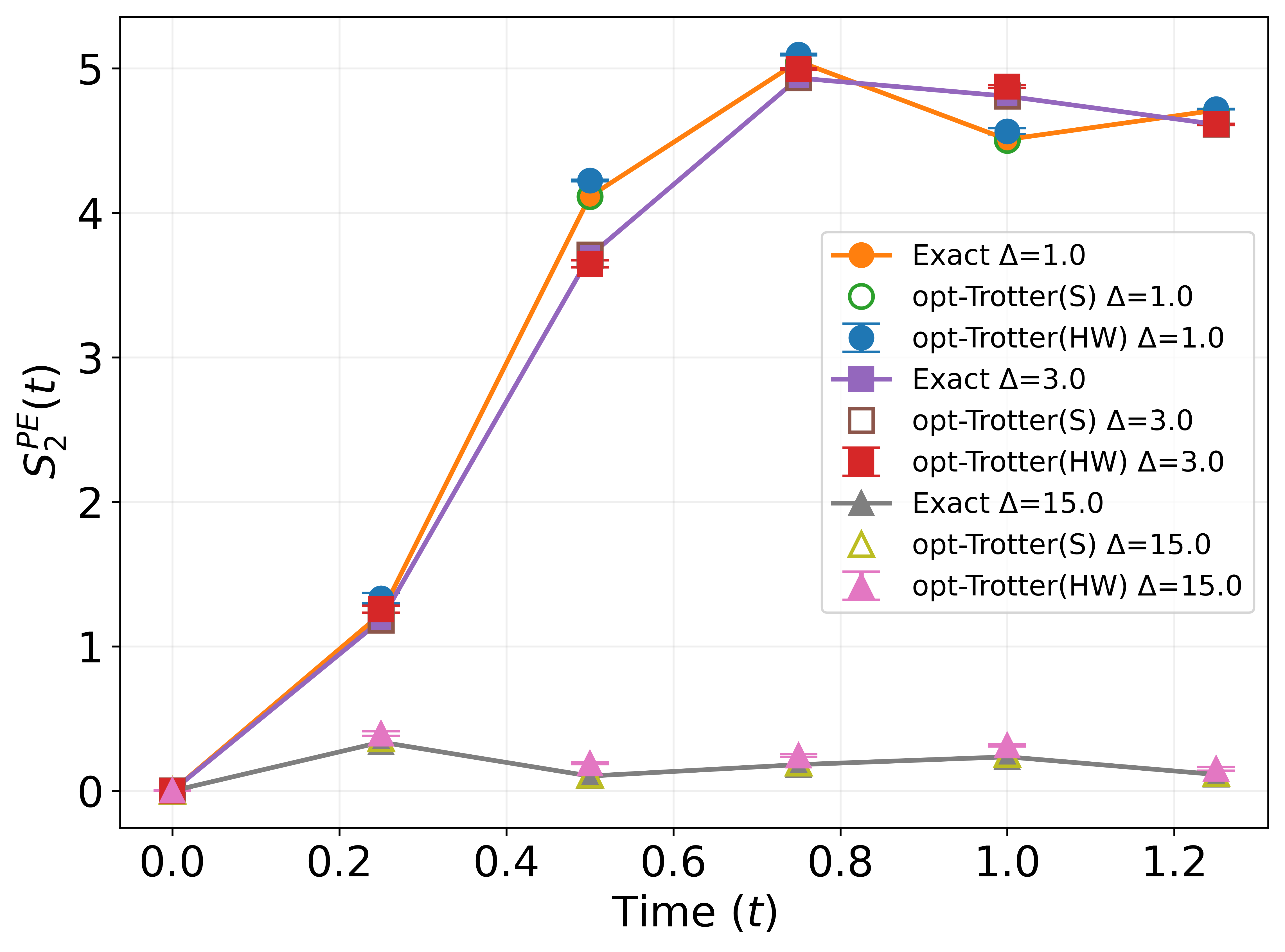}
\caption{Time evolution of the second-order participation entropy $S_2^{\mathrm{PE}}(t)$ the CDW initial state. Results are compared among exact diagonalization (Exact), optimized noiseless Trotter simulation [opt-Trotter(S)], and optimized Trotter hardware execution [opt-Trotter(HW)] for three Stark-field strengths, $\Delta_0=1$, $3$, and $15$. 
}
    \label{fig:s2_entropy}
\end{figure}

\begin{figure*}[tbh]
\centering
\subfigure[]{\includegraphics[width=0.49\textwidth]{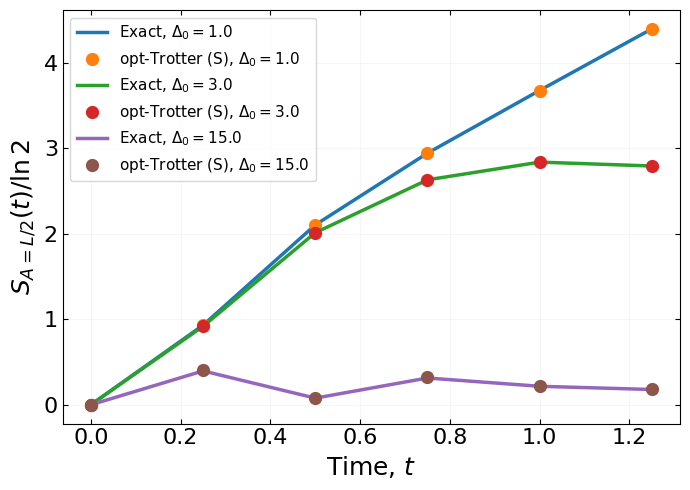}}
\subfigure[]{\includegraphics[width=0.49\textwidth]{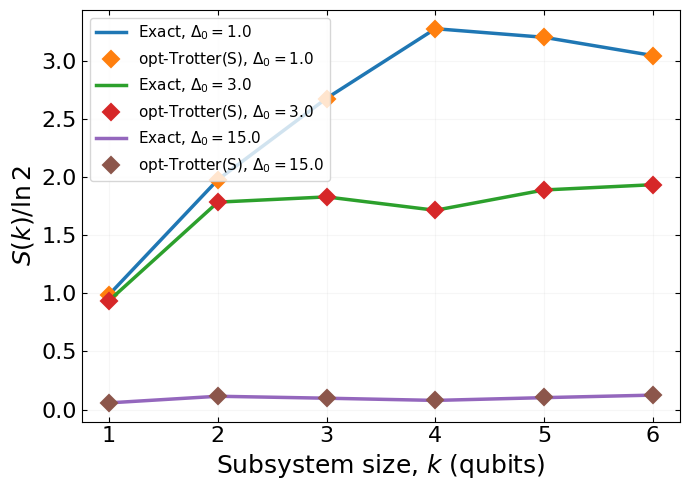}}
\caption{
von Neumann entanglement entropy for the CDW initial state obtained from exact diagonalization and optimized first-order Trotter simulation. (a) Half-chain entanglement entropy $S_{A=L/2}(t)$ as a function of time for three Stark-field strengths, $\Delta_0=1$, $\Delta_0=3$, and $\Delta_0=15$. Weak and intermediate Stark fields exhibit substantial entanglement growth, whereas strong Stark fields strongly suppress the buildup of entanglement. (b) Subsystem entanglement entropy $S(k)$ evaluated at the final evolution time $t=1.25$ as a function of subsystem size $k$. For weak tilts, the entropy increases rapidly with subsystem size, reflecting extensive information spreading. In contrast, the strong-tilt regime exhibits uniformly low entanglement across all subsystem sizes, consistent with Stark-MBL. The excellent agreement between exact diagonalization \cite{weinberg2017quspin} and optimized Trotter simulation confirms the accuracy of the quantum-circuit implementation.
}
\label{fig:entropy_halfchain_subsytemwise}
\end{figure*} 

\subsection{Circuit Optimization results}

Circuit depth and the number of two-qubit gates are key indicators of the practicality of executing quantum circuits on present-day hardware. Since two-qubit gates are significantly noisier than single-qubit operations and deeper circuits are more susceptible to decoherence, reducing both quantities is essential for improving the accuracy of hardware simulations. Before hardware execution, the logical circuits are expressed using CX gates, which are automatically transpiled to the native CZ gates of the IBM Heron2 architecture.

The effectiveness of the proposed optimization workflow is summarized in Table~\ref{tab:circuit_gate_counts_depth}. The table compares the original Trotter circuits with the optimized circuits obtained after applying AQC-Tensor, followed by the Qiskit L3--pytket-Qiskit L3 transpilation sequence, for both the CDW and DW initial states.
As expected, the size of the unoptimized Trotter circuit increases almost linearly with the evolution time since each additional Trotter step contributes another layer of gates.
For the $12$-qubit circuits considered in this work, a single unoptimized Trotter layer contains approximately $253$ two-qubit gates with a circuit depth of $213$. Therefore, the circuit size grows approximately linearly with the evolution time, reaching about $1265$ two-qubit gates and a circuit depth exceeding $1000$ at $t=1.25$. Such rapid growth in circuit complexity significantly increases hardware noise and motivates the circuit compression strategy adopted in this work.

In contrast, the optimized circuits remain nearly constant in size after compression. For example, while the unoptimized circuit grows from 253 to 1261 two-qubit gates as the evolution time increases from $t=0.25$ to $1.25$, the optimized circuits typically require only $150$--$235$ two-qubit gates, depending on the tilt strength and initial state.
The reduction becomes increasingly significant at longer evolution times. For the weakest tilt ($\Delta_{0}=1$), the two-qubit gate count is reduced by more than $81\%$ at $t=1.25$, with a corresponding circuit-depth reduction of approximately $80\%$. Even larger improvements are obtained for the strong-field regime ($\Delta_{0}=15$), where the optimized circuits achieve up to $87.9\%$ reduction in two-qubit gates and about $86.6\%$ reduction in circuit depth. Similar compression ratios are observed for both the CDW and DW initial states, indicating that the optimization is largely independent of the chosen initial state.
The resulting shallow circuits are therefore considerably better suited for execution on present-day superconducting quantum processors.

\begin{figure*}[tbh]
\centering
\subfigure[]{\includegraphics[width=0.3\textwidth]{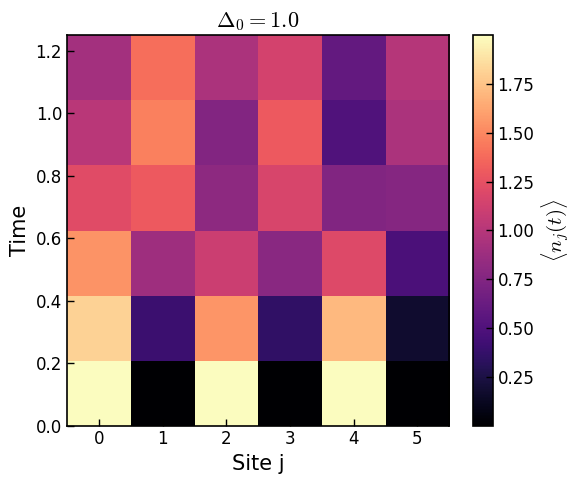}}
\subfigure[]{\includegraphics[width=0.3\textwidth]{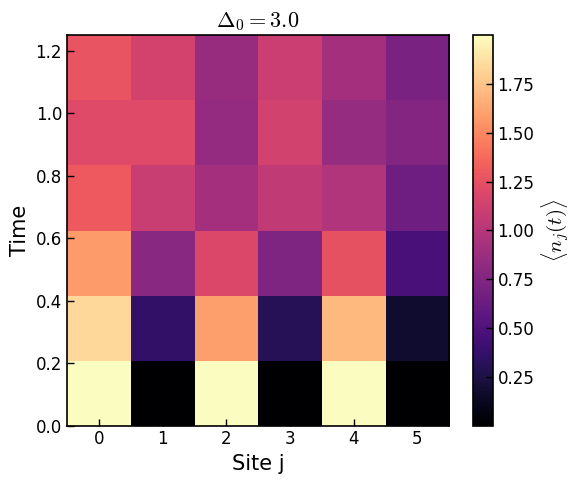}}
\subfigure[]{\includegraphics[width=0.3\textwidth]{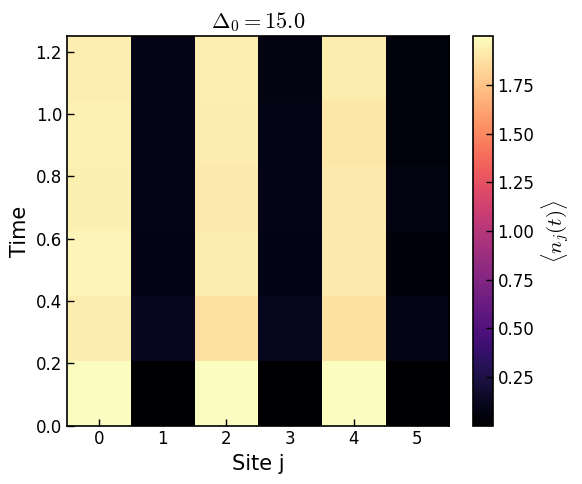}}
\subfigure[]{\includegraphics[width=0.31\textwidth]{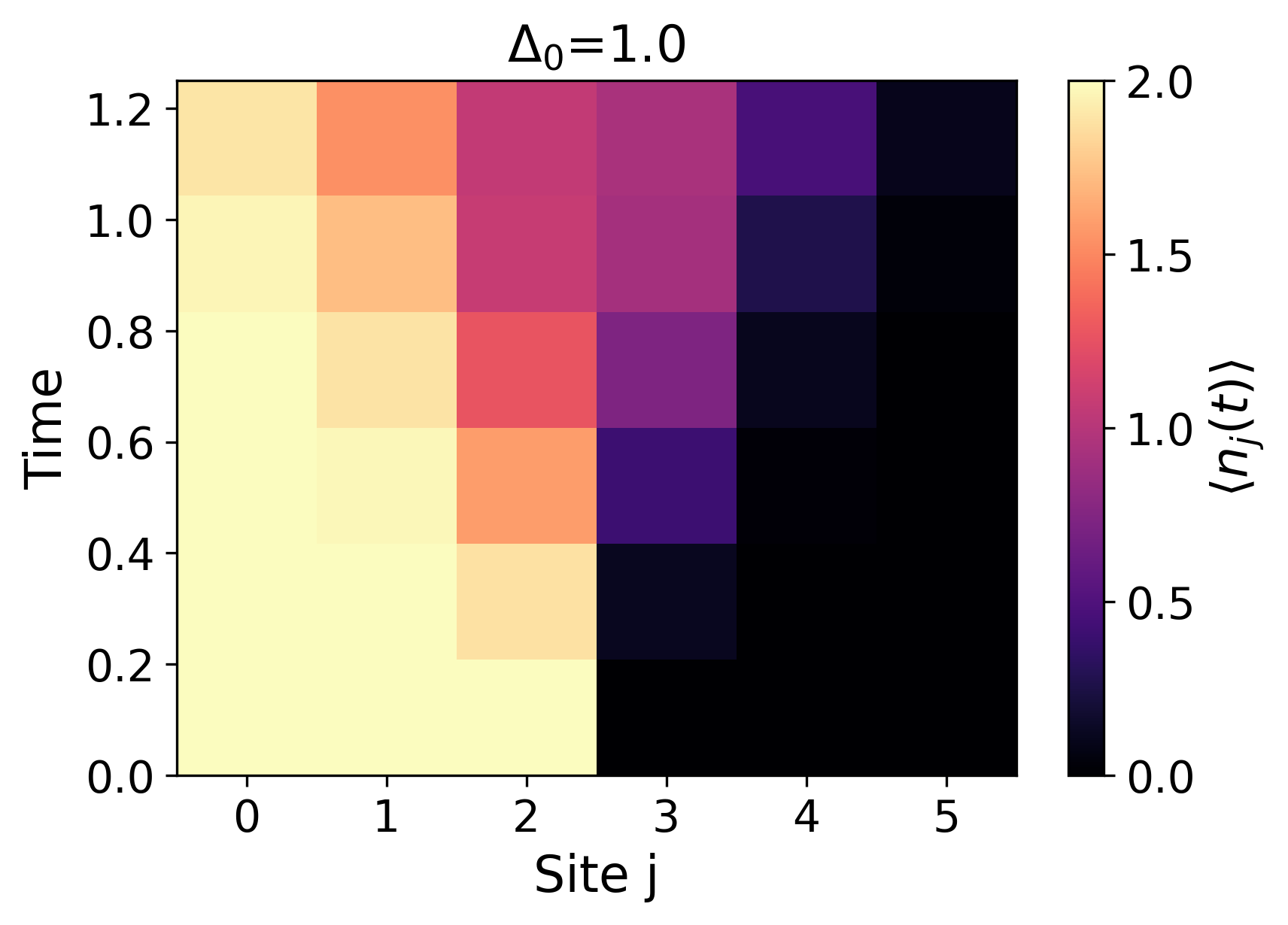}}
\subfigure[]{\includegraphics[width=0.31\textwidth]{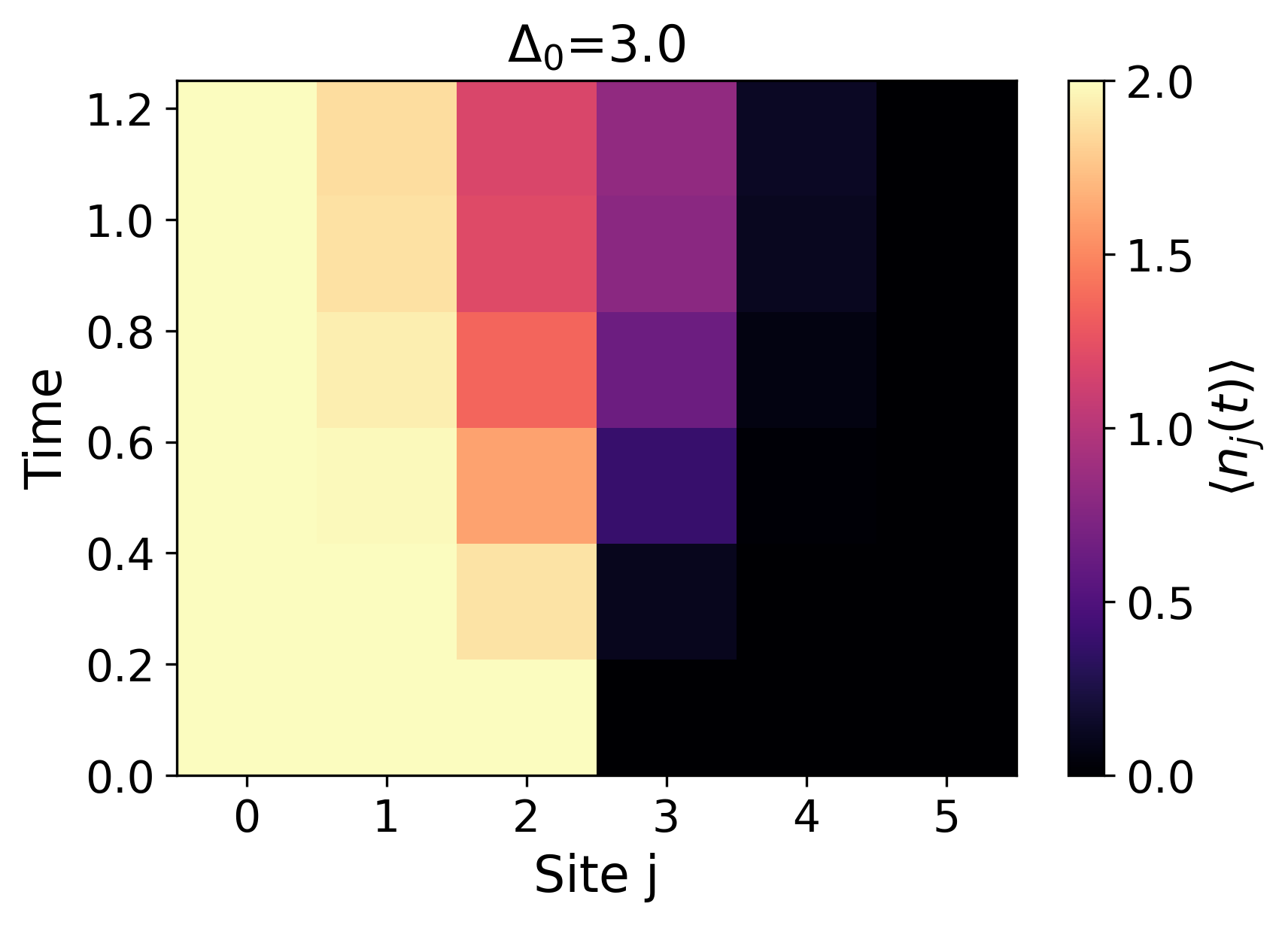}}
\subfigure[]{\includegraphics[width=0.31\textwidth]{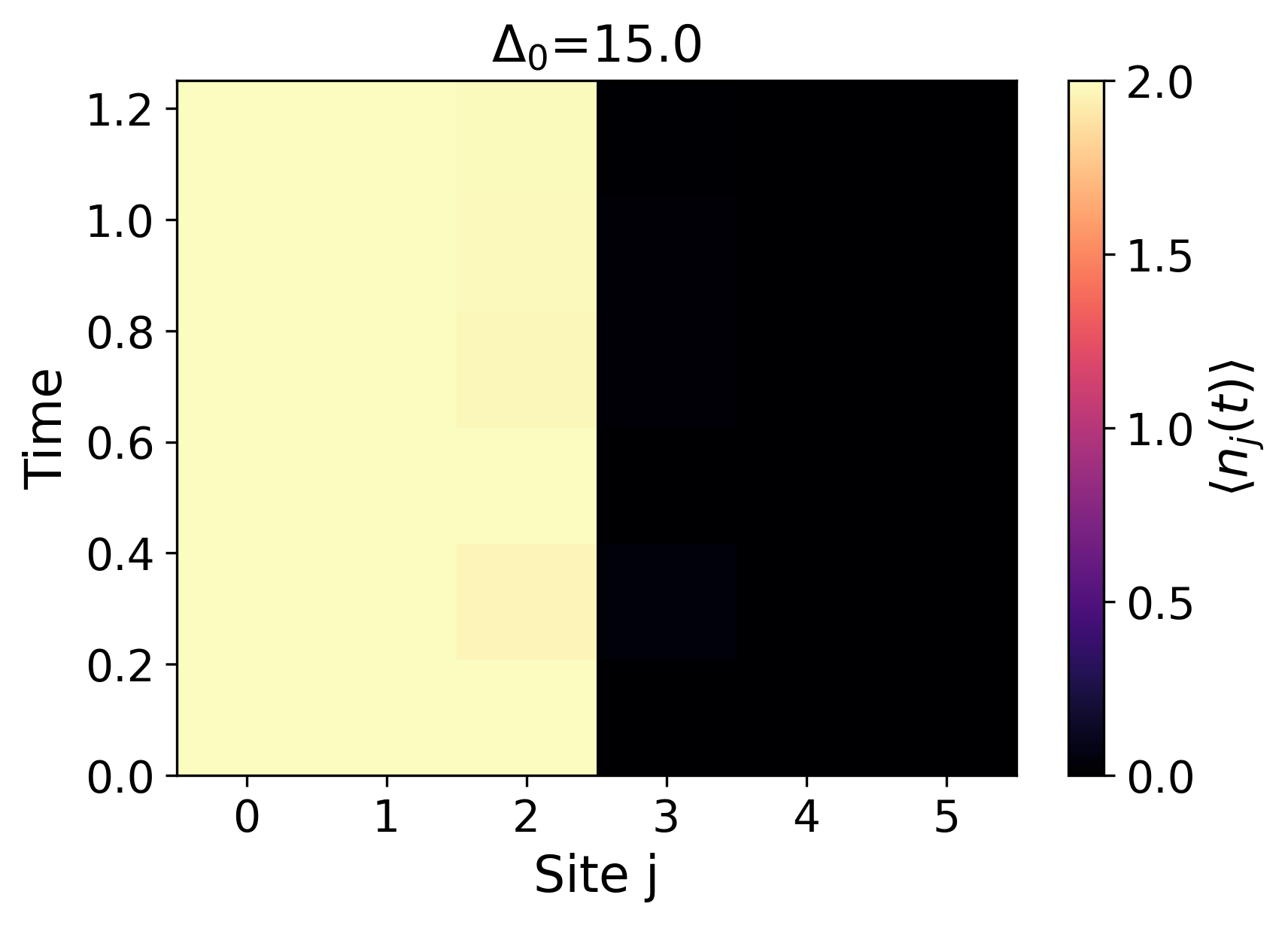}}
\caption{
Time evolution of the local particle density $\langle n_j(t)\rangle$ obtained from quantum-hardware experiments for different Stark-field strengths. Panels (a)-(c) correspond to the CDW initial state, while panels (d)-(f) correspond to the DW initial state. Results are shown for $\Delta_0=1$, $\Delta_0=3$, and $\Delta_0=15$, respectively. The color scale represents the site-resolved particle density as a function of time. For weak and intermediate tilts, particles gradually spread across the lattice, indicating finite transport and delocalization. In contrast, for the strong-field regime ($\Delta_0=15$), the density profile remains close to its initial configuration throughout the evolution, demonstrating strongly suppressed particle transport and the emergence of Stark-MBL. All data were obtained on the IBM \textit{ibm\_marrakesh} quantum processor after particle-number postselection and error-suppression protocols.
}
\label{fig:density_evolution}
\end{figure*}

\begin{figure}[tbh]
    \centering
    \includegraphics[width=0.95\linewidth]{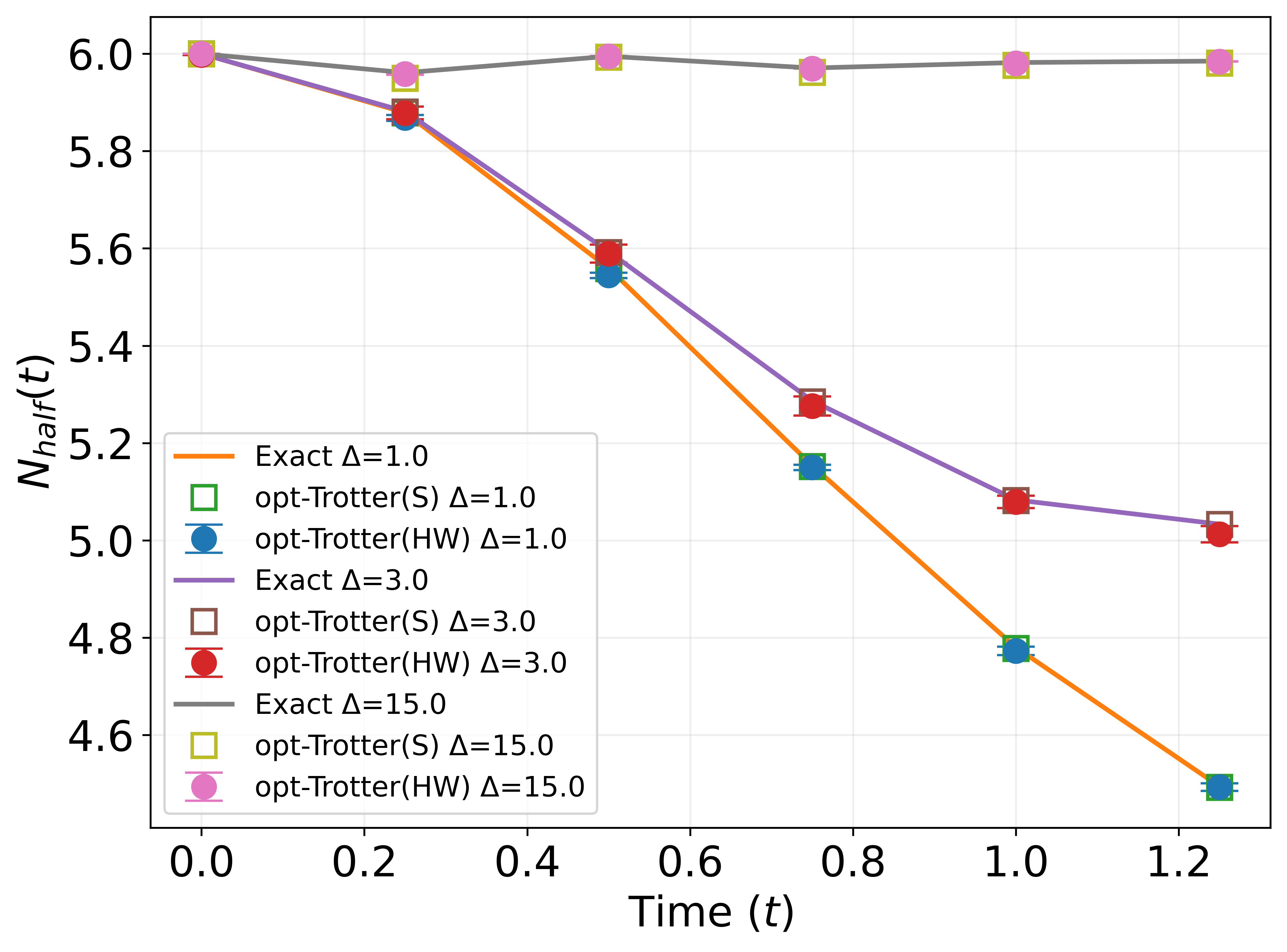}
\caption{
Time evolution of the particle number in the initially occupied half of the chain, $N_{\mathrm{half}}(t)$, for the DW initial state. Results are compared among exact diagonalization (Exact), optimized noiseless Trotter simulation [opt-Trotter(S)], and optimized Trotter hardware execution [opt-Trotter(HW)] for three Stark-field strengths, $\Delta_0=1$, $3$, and $15$. For weak and intermediate tilts, $N_{\mathrm{half}}(t)$ decreases with time, showing that particles leave the initially occupied region and spread across the chain. In contrast, for the strong-field case $\Delta_0=15$, $N_{\mathrm{half}}(t)$ remains close to its initial value of six particles, demonstrating strong suppression of transport and retention of particles within their original half of the lattice. The hardware results reproduce the same localization trend observed in the exact and optimized Trotter simulations.
}
    \label{fig:N_half}
\end{figure}

\subsection{Charge Imbalance, Quantum Fisher Information, and Participation Entropy}
We first analyze the dynamics starting from the CDW initial state.
% Throughout the paper, we consider $J=U=1$.
% This state is particularly useful for probing Stark-MBL because it contains a strong initial density modulation. If the system thermalizes, hopping processes gradually wash out this modulation. If the Stark field dominates, the initial pattern remains partially frozen. We therefore examine three complementary observables: the charge imbalance, the Quantum Fisher Information (QFI), and the second-order participation entropy. 
Figure~\ref{fig:imbalance_qfi_s2} compares exact diagonalization, optimized noiseless Trotter simulation, and optimized hardware results for $\Delta_0=1$, $3$, and $15$. The imbalance dynamics in Figure~\ref{fig:imbalance_qfi_s2}(a) show the clearest signature of localization. For weak and intermediate tilts, $\Delta_0=1$ and $\Delta_0=3$, the imbalance rapidly decreases from its initial value. In this regime, the energy mismatch between neighboring lattice sites is not large enough to prevent hopping, and particles can move away from their initially occupied sites. As a result, the CDW pattern and the system progressively lose memory of the initial state.

The behavior changes qualitatively for the strong tilt $\Delta_0=15$. Here, the linear potential creates a large energy gap between neighboring sites, making hopping processes unfavorable. Consequently, particles remain confined close to their initial locations, and the imbalance stays near its initial value over the accessible time window. 

The QFI shown in Figure~\ref{fig:imbalance_qfi_s2}(b) gives complementary information about the growth of many-body correlations. For $\Delta_0=1$ and $\Delta_0=3$, the rapid increase of $F_Q$ indicates that the initially simple product state develops nontrivial correlations during the evolution. This is expected when particles are able to move through the lattice: hopping and interactions together generate entanglement and multipartite correlations. In contrast, for $\Delta_0=15$, the QFI remains much smaller. The strong Stark field restricts particle motion and therefore limits the spreading of correlations. Thus, the slow QFI growth supports the localized interpretation obtained from the imbalance.

The participation entropy in Figure~\ref{fig:s2_entropy} probes the same dynamics from the viewpoint of Hilbert-space exploration. In the weak- and intermediate-field regimes, $S_2^{\mathrm{PE}}$ grows rapidly, showing that the wavefunction spreads over many many-body basis states. This behavior is consistent with delocalized dynamics, where the initial CDW state evolves into a superposition of many configurations. For $\Delta_0=15$, however, $S_2^{\mathrm{PE}}$ remains small, indicating that the dynamics involve only a limited part of the Hilbert space. 
% The system therefore remains close to the initial configuration not only in \textcolor{red}{real space}, as seen from the imbalance, but also in the many-body basis.

The three observables therefore give a consistent physical picture. The optimized noiseless Trotter data closely follow the exact results, confirming that the compressed circuits reproduce the target dynamics. The hardware measurements show the same qualitative trends, demonstrating that the optimized circuits, together with error suppression and postselection, capture the finite-time signatures of Stark localization.

To obtain a more detailed picture of the spatial structure of entanglement, we also evaluate the subsystem entropy $S(k)$ as a function of subsystem size $k$ at the final evolution time, as shown in Figure~\ref{fig:entropy_halfchain_subsytemwise}(b). For $\Delta_0=1$, the entropy increases rapidly with subsystem size and reaches its maximum near the half-chain partition, reflecting extensive correlation spreading across the lattice. The intermediate-field case $\Delta_0=3$ exhibits a similar but reduced trend.
In contrast, for $\Delta_0=15$, the subsystem entropy remains close to zero for all subsystem sizes. The weak dependence on subsystem size indicates that quantum correlations remain localized and do not spread significantly beyond their initial region. Such behavior is a characteristic signature of localized dynamics and provides further evidence for the formation of a Stark-MBL phase.

\subsection{Density Evolution and Suppression of Particle Transport}

A direct signature of Stark localization is obtained from the time evolution of the local density profile. Figure~\ref{fig:density_evolution} shows the site-resolved density $\langle n_j(t)\rangle$ measured on quantum hardware for both the CDW and DW initial states at different Stark-field strengths.

The upper row [Figures~\ref{fig:density_evolution}(a)--(c)] corresponds to the CDW initial state. At $t=0$, the particles occupy alternating lattice sites, producing a strong density modulation. For weak tilt, $\Delta_0=1$, this pattern rapidly becomes distorted as particles hop away from their initially occupied sites. The density profile becomes more uniform with time, indicating particle transport and loss of memory of the initial CDW configuration. A similar trend is observed for the intermediate tilt, $\Delta_0=3$, although the spreading is slower due to the larger energy gradient.
For the strong-field case, $\Delta_0=15$, the behavior changes qualitatively. The density modulation remains largely intact over the accessible evolution time, showing that particles stay close to their initial positions which is consistent with the large imbalance observed in Figure~\ref{fig:imbalance_qfi_s2}.

The lower row [Figures~\ref{fig:density_evolution}(d)--(f)] shows the corresponding density evolution for the DW initial state. In this case, particles initially occupy only one half of the lattice. For $\Delta_0=1$, the domain wall broadens as particles propagate into the initially empty region. The intermediate tilt again shows reduced spreading, indicating partial suppression of transport. In contrast, for $\Delta_0=15$, the domain wall remains almost frozen: the particles remain confined to the initially occupied half, with only weak density fluctuations near the interface.
To quantify this behavior, we compute the particle number in the initially occupied half of the chain, $N_{\mathrm{half}}(t)$, defined in Eq.~(\ref{eq:Nhalf}). For the 12-qubit system considered here, the DW state starts with $N_{\mathrm{half}}(0)=6$. As particles cross the center of the chain, $N_{\mathrm{half}}(t)$ decreases. Figure~\ref{fig:N_half} shows that this decrease is substantial for $\Delta_0=1$ and remains visible for $\Delta_0=3$, confirming transport across the domain wall. By contrast, for $\Delta_0=15$, $N_{\mathrm{half}}(t)$ stays close to its initial value, demonstrating that nearly all particles remain trapped in their original half of the lattice.

The density heatmaps and $N_{\mathrm{half}}(t)$ therefore give a consistent picture of the dynamics. Weak and intermediate Stark fields allow particles to redistribute across the chain, whereas a strong Stark field suppresses transport and preserves the initial density structure. These observations provide direct evidence for the emergence of Stark-localized dynamics in the large-tilt regime.
The \textit{percentage fractional differences} (PFDs) between the quantum-hardware and exact-diagonalization results for the observables considered here are provided in the Supplemental Material appendix C.

Although Trotterized evolution provides a systematic approach to simulating many-body dynamics, accessing long evolution times in present-day quantum hardware remains challenging. We recall that $\Delta t$ equals $t/n$. For a fixed time step $\Delta t$, the number of Trotter steps grows linearly with the total evolution time, thus leading to deep circuits when the required $t$ is long. On the other hand, for a fixed but sufficiently long evolution time, reducing $\Delta t$ to suppress Trotter errors further increases the circuit depth. Consequently, both the number of two-qubit gates and the accumulated hardware noise increase rapidly in practice with simulation time on NISQ computers. For a system of $\mathcal{N}$ qubits, let $m_{2q}(\mathcal{N})$ denote the number of two-qubit gates required for a single Trotter step and $d_{\mathrm{step}}(\mathcal{N})$ the corresponding circuit depth. For a total evolution time $t$ and Trotter step size $\Delta t$, the number of Trotter steps is $n=\frac{t}{\Delta t}$. Therefore, the total two-qubit gate count and total circuit depth scale as $G_{2q}(t,\mathcal{N})
=m_{2q}(\mathcal{N})
\frac{t}{\Delta t}$ and $D(t,\mathcal{N}) =d_{\mathrm{step}}(\mathcal{N})
\frac{t}{\Delta t}.$
This scaling shows why accessing long-time dynamics is difficult on present-day hardware. Even when the depth of a single Trotter layer is optimized, the total number of noisy two-qubit gates grows linearly with the simulated time. If the average two-qubit gate error is denoted by $\epsilon_{2q}$, the idealized circuit success probability scales approximately as $P_{\mathrm{succ}}
\sim
(1-\epsilon_{2q})^{G_{2q}}
\approx
e^{-\epsilon_{2q}G_{2q}}.$
Thus, for long-time simulations, even a small two-qubit gate error can lead to a rapid loss of circuit fidelity.

\section{Conclusion}\label{sec:4}
In this work, we have studied Stark MBL in a correlated fermionic system using Hamiltonian simulation on an IBM quantum computer. We have considered the tilted Fermi-Hubbard model for this purpose. To implement the dynamics efficiently on NISQ hardware, we employed a spin-resolved Jordan--Wigner mapping together with a SWAP network. The mapping reduces the two-qubit gate cost of the hopping blocks, while the SWAP network brings the required next-nearest-neighbor qubits into adjacent positions. As the resulting circuit is constructed for linear qubit connectivity, it also avoids the need for additional routing operations during hardware implementation. We have further employed a circuit-compression pipeline combining tensor network-based circuit optimization like AQC-Tensor with the standard optimization routines available in Qiskit and pytket. The optimized circuits achieve reductions of up to approximately $88\%$ in two-qubit gate count and $87\%$ in circuit depth relative to the original Trotterized circuits.
The dynamics has been characterized through complementary observables, including charge imbalance, site-resolved particle density, participation entropy, quantum Fisher information, and half-chain particle number. Their behavior at different Stark-field strengths is verified on a quantum computer, which demonstrates the suppression of particle transport and the memory retention of the initial state in the strongly tilted regime.
Our results show that present-day quantum hardware can be used to study dynamical phenomena like MBL in correlated quantum many-body systems, provided simulations of the systems on these devices are resource-efficient.

\section{Acknowledgment}
AK would like to acknowledge Dr. Nilanjan Roy for useful discussions on many-body localization and the Fermi-Hubbard model.

\bibliographystyle{plain}
\bibliography{sample}

@article{chowdhury2026quantum,
  title={Quantum utility in simulating the real-time dynamics of the Fermi--Hubbard model using superconducting quantum computers},
  author={Chowdhury, Talal Ahmed and Korepin, Vladimir and Pascuzzi, Vincent R and Yu, Kwangmin},
  journal={Applied Physics Reviews},
  volume={13},
  number={1},
  year={2026},
  publisher={AIP Publishing}
}

@article{bassman2022constant,
  title={Constant-depth circuits for dynamic simulations of materials on quantum computers},
  author={Bassman Oftelie, Lindsay and Van Beeumen, Roel and Younis, Ed and Smith, Ethan and Iancu, Costin and de Jong, Wibe A},
  journal={Materials Theory},
  volume={6},
  number={1},
  pages={13},
  year={2022},
  publisher={Springer}
}

@article{anselmetti2021local,
  title={Local, expressive, quantum-number-preserving VQE ans{\"a}tze for fermionic systems},
  author={Anselmetti, Gian-Luca R and Wierichs, David and Gogolin, Christian and Parrish, Robert M},
  journal={New Journal of Physics},
  volume={23},
  number={11},
  pages={113010},
  year={2021},
  publisher={IOP Publishing}
}

@article{zhang2024optimal,
  title={Optimal realization of yang--baxter gate on quantum computers},
  author={Zhang, Kun and Yu, Kwangmin and Hao, Kun and Korepin, Vladimir},
  journal={Advanced Quantum Technologies},
  volume={7},
  number={4},
  pages={2300345},
  year={2024},
  publisher={Wiley Online Library}
}

@article{jordan1928paulische,
  title={{\"U}ber das paulische {\"a}quivalenzverbot},
  author={Jordan, Pascual and Wigner, Eugene},
  journal={Zeitschrift f{\"u}r Physik},
  volume={47},
  number={9},
  pages={631--651},
  year={1928},
  publisher={Springer}
}

@article{trotter1959product,
  title={On the product of semi-groups of operators},
  author={Trotter, Hale F},
  journal={Proceedings of the American Mathematical Society},
  volume={10},
  number={4},
  pages={545--551},
  year={1959},
  publisher={JSTOR}
}

@article{suzuki1976generalized,
  title={Generalized Trotter's formula and systematic approximants of exponential operators and inner derivations with applications to many-body problems},
  author={Suzuki, Masuo},
  journal={Communications in Mathematical Physics},
  volume={51},
  number={2},
  pages={183--190},
  year={1976},
  publisher={Springer}
}

@article{chowdhury2024enhancing,
  title={Enhancing quantum utility: Simulating large-scale quantum spin chains on superconducting quantum computers},
  author={Chowdhury, Talal Ahmed and Yu, Kwangmin and Shamim, Mahmud Ashraf and Kabir, ML and Sufian, Raza Sabbir},
  journal={Physical Review Research},
  volume={6},
  number={3},
  pages={033107},
  year={2024},
  publisher={APS}
}

@article{weinberg2017quspin,
  title={QuSpin: a Python package for dynamics and exact diagonalisation of quantum many body systems part I: spin chains},
  author={Weinberg, Phillip and Bukov, Marin},
  journal={SciPost Physics},
  volume={2},
  number={1},
  pages={003},
  year={2017}
}

@article{robertson2025approximate,
  title={Approximate quantum compiling for quantum simulation: A tensor network based approach},
  author={Robertson, Niall and Akhriev, Albert and Vala, Jiri and Zhuk, Sergiy},
  journal={ACM Transactions on Quantum Computing},
  volume={6},
  number={3},
  pages={1--15},
  year={2025},
  publisher={ACM New York, NY}
}

@article{bennett1996purification,
  title={Purification of noisy entanglement and faithful teleportation via noisy channels},
  author={Bennett, Charles H and Brassard, Gilles and Popescu, Sandu and Schumacher, Benjamin and Smolin, John A and Wootters, William K},
  journal={Physical Review Letters},
  volume={76},
  number={5},
  pages={722},
  year={1996},
  publisher={APS}
}

@article{wallman2016noise,
  title={Noise tailoring for scalable quantum computation via randomized compiling},
  author={Wallman, Joel J and Emerson, Joseph},
  journal={Physical Review A},
  volume={94},
  number={5},
  pages={052325},
  year={2016},
  publisher={APS}
}

@article{cai2019constructing,
  title={Constructing smaller Pauli twirling sets for arbitrary error channels},
  author={Cai, Zhenyu and Benjamin, Simon C},
  journal={Scientific Reports},
  volume={9},
  number={1},
  pages={11281},
  year={2019},
  publisher={Nature Publishing Group UK London}
}

@article{preskill2018quantum,
  title={Quantum computing in the NISQ era and beyond},
  author={Preskill, John},
  journal={Quantum},
  volume={2},
  pages={79},
  year={2018},
  publisher={Verein zur F{\"o}rderung des Open Access Publizierens in den Quantenwissenschaften}
}

@article{georgescu2014quantum,
  title={Quantum simulation},
  author={Georgescu, Iulia M and Ashhab, Sahel and Nori, Franco},
  journal={Reviews of Modern Physics},
  volume={86},
  number={1},
  pages={153--185},
  year={2014},
  publisher={APS}
}

@article{sivarajah2020t,
  title={t| ket>: a retargetable compiler for NISQ devices},
  author={Sivarajah, Seyon and Dilkes, Silas and Cowtan, Alexander and Simmons, Will and Edgington, Alec and Duncan, Ross},
  journal={Quantum Science and Technology},
  volume={6},
  number={1},
  pages={014003},
  year={2020},
  publisher={IOP Publishing}
}

@article{javadi2024quantum,
  title={Quantum computing with Qiskit},
  author={Javadi-Abhari, Ali and Treinish, Matthew and Krsulich, Kevin and Wood, Christopher J and Lishman, Jake and Gacon, Julien and Martiel, Simon and Nation, Paul D and Bishop, Lev S and Cross, Andrew W and R. Johnson, Blake and Gambetta , Jay M.},
  journal={arXiv preprint arXiv:2405.08810},
  year={2024}
}

@article{viola1999dynamical,
  title={Dynamical decoupling of open quantum systems},
  author={Viola, Lorenza and Knill, Emanuel and Lloyd, Seth},
  journal={Physical Review Letters},
  volume={82},
  number={12},
  pages={2417},
  year={1999},
  publisher={APS}
}

@article{ezzell2022dynamical,
    title = "{Dynamical decoupling for superconducting qubits: A performance survey}",
    author = "Ezzell, Nic and Pokharel, Bibek and Tewala, Lina and Quiroz, Gregory and Lidar, Daniel A.",
    journal = "Phys. Rev. Applied",
    volume = "20",
    number = "6",
    pages = "064027",
    year = "2023"
}

@article{liu1989limited,
  title={On the limited memory BFGS method for large scale optimization},
  author={Liu, Dong C and Nocedal, Jorge},
  journal={Mathematical Programming},
  volume={45},
  number={1},
  pages={503--528},
  year={1989},
  publisher={Springer}
}

@article{guo2021stark,
  title={Stark many-body localization on a superconducting quantum processor},
  author={Guo, Qiujiang and Cheng, Chen and Li, Hekang and Xu, Shibo and Zhang, Pengfei and Wang, Zhen and Song, Chao and Liu, Wuxin and Ren, Wenhui and Dong, Hang and Mondaini , Rubem and  Wang, H.},
  journal={Physical Review Letters},
  volume={127},
  number={24},
  pages={240502},
  year={2021},
  publisher={APS}
}

@article{battistel2023real,
  title={Real-time decoding for fault-tolerant quantum computing: Progress, challenges and outlook},
  author={Battistel, Francesco and Chamberland, Christopher and Johar, Kauser and Overwater, Ramon WJ and Sebastiano, Fabio and Skoric, Luka and Ueno, Yosuke and Usman, Muhammad},
  journal={Nano Futures},
  volume={7},
  number={3},
  pages={032003},
  year={2023},
  publisher={IOP Publishing}
}

@article{fauseweh2024quantum,
  title={Quantum many-body simulations on digital quantum computers: State-of-the-art and future challenges},
  author={Fauseweh, Benedikt},
  journal={Nature Communications},
  volume={15},
  number={1},
  pages={2123},
  year={2024},
  publisher={Nature Publishing Group UK London}
}

@article{deutsch1991quantum,
  title={Quantum statistical mechanics in a closed system},
  author={Deutsch, Josh M},
  journal={Physical Review A},
  volume={43},
  number={4},
  pages={2046},
  year={1991},
  publisher={APS}
}

@article{srednicki1994chaos,
  title={Chaos and quantum thermalization},
  author={Srednicki, Mark},
  journal={Physical Review E},
  volume={50},
  number={2},
  pages={888},
  year={1994},
  publisher={APS}
}

@article{nandkishore2015many,
  title={Many-body localization and thermalization in quantum statistical mechanics},
  author={Nandkishore, Rahul and Huse, David A},
  journal={Annual Review of Condensed Matter Physics},
  volume={6},
  number={1},
  pages={15--38},
  year={2015},
  publisher={Annual Reviews}
}

@article{anderson1958absence,
  title={Absence of diffusion in certain random lattices},
  author={Anderson, Philip W},
  journal={Physical Review},
  volume={109},
  number={5},
  pages={1492--1505},
  year={1958}
}

@article{zhu2021probing,
  title={Probing many-body localization on a noisy quantum computer},
  author={Zhu, Daiwei and Johri, S and Nguyen, NH and Alderete, C Huerta and Landsman, KA and Linke, NM and Monroe, C and Matsuura, AY},
  journal={Physical Review A},
  volume={103},
  number={3},
  pages={032606},
  year={2021},
  publisher={APS}
}

@article{schreiber2015observation,
  title={Observation of many-body localization of interacting fermions in a quasirandom optical lattice},
  author={Schreiber, Michael and Hodgman, Sean S and Bordia, Pranjal and L{\"u}schen, Henrik P and Fischer, Mark H and Vosk, Ronen and Altman, Ehud and Schneider, Ulrich and Bloch, Immanuel},
  journal={Science},
  volume={349},
  number={6250},
  pages={842--845},
  year={2015},
  publisher={American Association for the Advancement of Science}
}

@article{hyllus2012fisher,
  title={Fisher information and multiparticle entanglement},
  author={Hyllus, Philipp and Laskowski, Wies{\l}aw and Krischek, Roland and Schwemmer, Christian and Wieczorek, Witlef and Weinfurter, Harald and Pezz{\'e}, Luca and Smerzi, Augusto},
  journal={Physical Review A},
  volume={85},
  number={2},
  pages={022321},
  year={2012},
  publisher={APS}
}

@article{toth2012multipartite,
  title={Multipartite entanglement and high-precision metrology},
  author={T{\'o}th, G{\'e}za},
  journal={Physical Review A},
  volume={85},
  number={2},
  pages={022322},
  year={2012},
  publisher={APS}
}

@article{smith2019simulating,
  title={Simulating quantum many-body dynamics on a current digital quantum computer},
  author={Smith, Adam and Kim, Myungshik S and Pollmann, Frank and Knolle, Johannes},
  journal={npj Quantum Information},
  volume={5},
  number={1},
  pages={106},
  year={2019},
  publisher={Nature Publishing Group UK London}
}

@article{arovas2022hubbard,
  title={The hubbard model},
  author={Arovas, Daniel P and Berg, Erez and Kivelson, Steven A and Raghu, Srinivas},
  journal={Annual Review of Condensed Matter Physics},
  volume={13},
  number={1},
  pages={239--274},
  year={2022},
  publisher={Annual Reviews}
}

@article{grosse1989symmetry,
  title={The symmetry of the Hubbard model},
  author={Grosse, Harald},
  journal={Letters in Mathematical Physics},
  volume={18},
  number={2},
  pages={151--156},
  year={1989},
  publisher={Springer}
}

@article{stanisic2022observing,
  title={Observing ground-state properties of the Fermi-Hubbard model using a scalable algorithm on a quantum computer},
  author={Stanisic, Stasja and Bosse, Jan Lukas and Gambetta, Filippo Maria and Santos, Raul A and Mruczkiewicz, Wojciech and O’Brien, Thomas E and Ostby, Eric and Montanaro, Ashley},
  journal={Nature Communications},
  volume={13},
  number={1},
  pages={5743},
  year={2022},
  publisher={Nature Publishing Group UK London}
}

@article{ying2023experimental,
  title={Experimental simulation of larger quantum circuits with fewer superconducting qubits},
  author={Ying, Chong and Cheng, Bin and Zhao, Youwei and Huang, He-Liang and Zhang, Yu-Ning and Gong, Ming and Wu, Yulin and Wang, Shiyu and Liang, Futian and Lin, Jin and Yu, Xu and Hui, Deng and Hao, Rong and Peng, Cheng-Zhi and Yung, Man-Hong and Zhu, Xiaobo and Pan, Jian-Wei},
  journal={Physical Review Letters},
  volume={130},
  number={11},
  pages={110601},
  year={2023},
  publisher={APS}
}

@article{kandala2017hardware,
  title={Hardware-efficient variational quantum eigensolver for small molecules and quantum magnets},
  author={Kandala, Abhinav and Mezzacapo, Antonio and Temme, Kristan and Takita, Maika and Brink, Markus and Chow, Jerry M and Gambetta, Jay M},
  journal={Nature},
  volume={549},
  number={7671},
  pages={242--246},
  year={2017},
  publisher={Nature Publishing Group}
}

@article{kandala2019error,
  title={Error mitigation extends the computational reach of a noisy quantum processor},
  author={Kandala, Abhinav and Temme, Kristan and C{\'o}rcoles, Antonio D and Mezzacapo, Antonio and Chow, Jerry M and Gambetta, Jay M},
  journal={Nature},
  volume={567},
  number={7749},
  pages={491--495},
  year={2019},
  publisher={Nature Publishing Group UK London}
}

@article{hubbard1963electron,
  title={Electron correlations in narrow energy bands},
  author={Hubbard, John},
  journal={Proceedings of the Royal Society of London. Series A. Mathematical and Physical Sciences},
  volume={276},
  number={1365},
  pages={238--257},
  year={1963},
  publisher={The Royal Society London}
}

@article{alet2018many,
  title={Many-body localization: An introduction and selected topics},
  author={Alet, Fabien and Laflorencie, Nicolas},
  journal={Comptes Rendus Physique},
  volume={19},
  number={6},
  pages={498--525},
  year={2018},
  publisher={Elsevier}
}

@article{abanin2019colloquium,
  title={Colloquium: Many-body localization, thermalization, and entanglement},
  author={Abanin, Dmitry A and Altman, Ehud and Bloch, Immanuel and Serbyn, Maksym},
  journal={Reviews of Modern Physics},
  volume={91},
  number={2},
  pages={021001},
  year={2019},
  publisher={APS}
}

@article{gopalakrishnan2020dynamics,
  title={Dynamics and transport at the threshold of many-body localization},
  author={Gopalakrishnan, Sarang and Parameswaran, SA},
  journal={Physics Reports},
  volume={862},
  pages={1--62},
  year={2020},
  publisher={Elsevier}
}

@article{sierant2025many,
  title={Many-body localization in the age of classical computing},
  author={Sierant, Piotr and Lewenstein, Maciej and Scardicchio, Antonello and Vidmar, Lev and Zakrzewski, Jakub},
  journal={Reports on Progress in Physics},
  volume={88},
  number={2},
  pages={026502},
  year={2025},
  publisher={IOP Publishing}
}

@article{zakrzewski2026many,
  title={Many-body localization},
  author={Zakrzewski, Jakub},
  journal={arXiv preprint arXiv:2604.12464},
  year={2026}
}

@article{wilkinson1996observation,
  title={Observation of atomic Wannier-Stark ladders in an accelerating optical potential},
  author={Wilkinson, SR and Bharucha, CF and Madison, KW and Niu, Qian and Raizen, MG},
  journal={Physical Review Letters},
  volume={76},
  number={24},
  pages={4512},
  year={1996},
  publisher={APS}
}

@article{taylor2020experimental,
  title={Experimental probes of Stark many-body localization},
  author={Taylor, Scott Richard and Schulz, Maximilian and Pollmann, Frank and Moessner, Roderich},
  journal={Physical Review B},
  volume={102},
  number={5},
  pages={054206},
  year={2020},
  publisher={APS}
}

@article{yao2020many,
  title={Many-body localization of bosons in an optical lattice: Dynamics in disorder-free potentials},
  author={Yao, Ruixiao and Zakrzewski, Jakub},
  journal={Physical Review B},
  volume={102},
  number={10},
  pages={104203},
  year={2020},
  publisher={APS}
}

@article{lang2022disorder,
  title={Disorder-free localization with Stark gauge protection},
  author={Lang, Haifeng and Hauke, Philipp and Knolle, Johannes and Grusdt, Fabian and Halimeh, Jad C},
  journal={Physical Review B},
  volume={106},
  number={17},
  pages={174305},
  year={2022},
  publisher={APS}
}

@article{wang2021stark,
  title={Stark many-body localization transitions in superconducting circuits},
  author={Wang, Yong-Yi and Sun, Zheng-Hang and Fan, Heng},
  journal={Physical Review B},
  volume={104},
  number={20},
  pages={205122},
  year={2021},
  publisher={APS}
}

@article{morong2021observation,
  title={Observation of Stark many-body localization without disorder},
  author={Morong, William and Liu, Fangli and Becker, Patrick and Collins, Kate S and Feng, Lei and Kyprianidis, Antonis and Pagano, Guido and You, Tianyu and Gorshkov, Alexey V and Monroe, Christopher},
  journal={Nature},
  volume={599},
  number={7885},
  pages={393--398},
  year={2021},
  publisher={Nature Publishing Group UK London}
}

@article{van2019bloch,
  title={From Bloch oscillations to many-body localization in clean interacting systems},
  author={van Nieuwenburg, Evert and Baum, Yuval and Refael, Gil},
  journal={Proceedings of the National Academy of Sciences},
  volume={116},
  number={19},
  pages={9269--9274},
  year={2019},
  publisher={National Academy of Sciences}
}

@article{schulz2019stark,
  title={Stark many-body localization},
  author={Schulz, Maximilian and Hooley, CA and Moessner, Roderich and Pollmann, F},
  journal={Physical Review Letters},
  volume={122},
  number={4},
  pages={040606},
  year={2019},
  publisher={APS}
}

@article{wannier1962dynamics,
  title={Dynamics of band electrons in electric and magnetic fields},
  author={Wannier, Gregory H},
  journal={Reviews of Modern Physics},
  volume={34},
  number={4},
  pages={645},
  year={1962},
  publisher={APS}
}

@article{zhang2021mobility,
  title={Mobility edge of Stark many-body localization},
  author={Zhang, Li and Ke, Yongguan and Liu, Wenjie and Lee, Chaohong},
  journal={Physical Review A},
  volume={103},
  number={2},
  pages={023323},
  year={2021},
  publisher={APS}
}

@article{scherg2021observing,
  title={Observing non-ergodicity due to kinetic constraints in tilted Fermi-Hubbard chains},
  author={Scherg, Sebastian and Kohlert, Thomas and Sala, Pablo and Pollmann, Frank and Hebbe Madhusudhana, Bharath and Bloch, Immanuel and Aidelsburger, Monika},
  journal={Nature Communications},
  volume={12},
  number={1},
  pages={4490},
  year={2021},
  publisher={Nature Publishing Group UK London}
}

@article{aubry1980analyticity,
  title={Analyticity breaking and Anderson localization in incommensurate lattices},
  author={Aubry, Serge and Andr{\'e}, Gilles},
  journal={Ann. Israel Phys. Soc},
  volume={3},
  number={133},
  pages={18},
  year={1980}
}

@article{lloyd1996universal,
  title={Universal quantum simulators},
  author={Lloyd, Seth},
  journal={Science},
  volume={273},
  number={5278},
  pages={1073--1078},
  year={1996},
  publisher={American Association for the Advancement of Science}
}

@article{alam2025fermionic,
  title={Fermionic dynamics on a trapped-ion quantum computer beyond exact classical simulation},
  author={Alam, Faisal and Bosse, Jan Lukas and {\v{C}}epait{\.e}, Ieva and Chapman, Adrian and Clinton, Laura and Crichigno, Marcos and Crosson, Elizabeth and Cubitt, Toby and Derby, Charles and Dowinton, Oliver and others},
  journal={arXiv preprint arXiv:2510.26300},
  year={2025}
}

@article{alam2026onset,
  title={Onset of ergodicity across scales on a digital quantum processor},
  author={Alam, Faisal and Crichigno, Marcos and Crosson, Elizabeth and Flammia, Steven T and Gambetta, Filippo Maria and Gordon, Max Hunter and Kreshchuk, Michael and Montanaro, Ashley and Nocera, Alberto and Santos, Raul A},
  journal={arXiv preprint arXiv:2603.12236},
  year={2026}
}

@article{cobos2025real,
  title={Real-time dynamics in a (2+ 1)-D gauge theory: The stringy nature on a superconducting quantum simulator},
  author={Cobos, Jes{\'u}s and Fraxanet, Joana and Benito, C{\'e}sar and di Marcantonio, Francesco and Rivero, Pedro and Kap{\'a}s, Korn{\'e}l and Werner, Mikl{\'o}s Antal and Legeza, {\"O}rs and Bermudez, Alejandro and Rico, Enrique},
  journal={arXiv preprint arXiv:2507.08088},
  year={2025}
}

@article{campbell2017roads,
  title={Roads towards fault-tolerant universal quantum computation},
  author={Campbell, Earl T and Terhal, Barbara M and Vuillot, Christophe},
  journal={Nature},
  volume={549},
  number={7671},
  pages={172--179},
  year={2017},
  publisher={Nature Publishing Group UK London}
}

\onecolumn\newpage
\appendix

\section{Derivation of the Quantum Fisher Information}

To characterize the growth of multipartite correlations in the Stark Fermi-Hubbard model, we evaluate the quantum Fisher information (QFI) associated with the staggered charge operator. The staggered operator is defined as

\begin{equation}
\hat{O}_c
=
\sum_j (-1)^j
\left(
n_{j\uparrow}
+
n_{j\downarrow}
\right),
\end{equation}
where $n_{j\uparrow}$ and $n_{j\downarrow}$ denote the local number operators for spin-up and spin-down fermions, respectively. The alternating factor $(-1)^j$ probes charge-density-wave order and captures the staggered nature of the initial state.
For a pure quantum state, the quantum Fisher information corresponding to the operator $\hat{O}_c$ is given by

\begin{equation}
F_Q
=
4
\left(
\langle \hat{O}_c^2 \rangle
-
\langle \hat{O}_c \rangle^2
\right),
\end{equation}
which is proportional to the variance of $\hat{O}_c$. A large value of $F_Q$ indicates the buildup of strong quantum correlations and multipartite entanglement during the dynamics.

To express the QFI in the qubit representation, we use the Jordan-Wigner mapping between fermionic number operators and Pauli-$Z$ operators,

\begin{equation}
n=\frac{1-\hat{\sigma}^z}{2}.
\end{equation}
For the spin-up and spin-down orbitals associated with a given lattice site $j$, the mapping becomes
\begin{equation}
n_{j\uparrow}=\frac{1-\hat{\sigma}^z_j}{2},
\qquad
n_{j\downarrow}=\frac{1-\hat{\sigma}^z_{j+1}}{2}.
\end{equation}
Substituting these relations into the staggered charge operator yields

\begin{equation}
n_{j\uparrow}
+
n_{j\downarrow}
=
\frac{
2
-
\hat{\sigma}^z_j
-
\hat{\sigma}^z_{j+1}
}{2}.
\end{equation}
Accordingly, the operator $\hat{O}_c$ can be rewritten as
\begin{equation}
\hat{O}_c
=
\sum_{j=0}^{L-1}
(-1)^j
\left(
\frac{
2
-
\hat{\sigma}^z_j
-
\hat{\sigma}^z_{j+1}
}{2}
\right).
\end{equation}
Separating the constant contribution from the operator-dependent terms gives
\begin{equation}
\hat{O}_c
=
\sum_j (-1)^j
-
\frac{1}{2}
\sum_j
(-1)^j
\left(
\hat{\sigma}^z_j
+
\hat{\sigma}^z_{j+1}
\right).
\end{equation}

Since constant terms do not contribute to the variance, they do not affect the quantum Fisher information and may therefore be omitted. The relevant operator entering the QFI is thus
\begin{equation}
\hat{O}_c
=
\frac{1}{2}
\sum_{j=0}^{L-1}
(-1)^j
\left(
\hat{\sigma}^z_j
+
\hat{\sigma}^z_{j+1}
\right).
\end{equation}

For compactness, we introduce coefficients $c_j$ defined by

\begin{equation}
c_j=
\begin{cases}
+1, & \text{for even sites}, \\
-1, & \text{for odd sites}.
\end{cases}
\end{equation}

The staggered operator can then be expressed in the form

\begin{equation}
\hat{O}_c
=
\frac{1}{2}
\sum_j
c_j
\hat{\sigma}^z_j.
\end{equation}

Squaring the operator gives

\begin{equation}
\hat{O}_c^2
=
\frac{1}{4}
\sum_{jk}
c_j c_k
\hat{\sigma}^z_j
\hat{\sigma}^z_k.
\end{equation}

Substituting this expression into the definition of the QFI finally yields

\begin{equation}
F_Q
=
\sum_{jk}
c_j c_k
\left(
\langle
\hat{\sigma}^z_j
\hat{\sigma}^z_k
\rangle
-
\langle
\hat{\sigma}^z_j
\rangle
\langle
\hat{\sigma}^z_k
\rangle
\right).
\end{equation}

Therefore, the quantum Fisher information is directly related to the connected two-point correlation functions of the Pauli-$Z$ operators. This form is particularly convenient for quantum simulations, since the required expectation values can be obtained from measurements in the computational basis.

% \section{The way we calculated on a quantum computer is the expectation value Calculation on a Quantum Computer}

\section{Evaluation of observables from quantum-hardware measurement outcomes}
\label{sec:hardware_observables}

The observables reported in this work are evaluated directly from the computational-basis measurement outcomes obtained from the quantum processor. For an \(L\)-site tilted Fermi--Hubbard system, \(2L\) qubits are used, with the spin orbitals mapped according to
\begin{equation}
(j,\uparrow)\rightarrow q_{2j},
\qquad
(j,\downarrow)\rightarrow q_{2j+1}.
\end{equation}
A measurement outcome is therefore represented by a bitstring

$$
b=(b_0,b_1,\ldots,b_{2L-1}),
\qquad b_q\in\{0,1\}.
$$

If a bitstring \(b\) is observed \(N_b\) times out of a total of \(N_{\mathrm{shots}}\) measurements, its probability is estimated as
\begin{equation}
p_b=\frac{N_b}{N_{\mathrm{shots}}},
\qquad
\sum_b p_b=1.
\label{eq:hardware_probability}
\end{equation}
For mitigated results, \(p_b\) is replaced by the corresponding mitigated probability distribution.

The fermionic occupation operator associated with qubit \(q\) is
\begin{equation}
\hat n_q=\frac{1-\hat Z_q}{2},
\label{eq:number_pauli}
\end{equation}
such that a measured bit \(b_q=0\) corresponds to \(n_q=0\), while \(b_q=1\) corresponds to \(n_q=1\). Equivalently, the eigenvalue of \(\hat Z_q\) for a given measurement outcome is
\begin{equation}
z_q(b)=(-1)^{b_q}.
\end{equation}
Therefore, expectation values of any operator diagonal in the computational basis can be obtained as
\begin{equation}
\langle \hat O\rangle
=
\sum_b p_b O(b),
\label{eq:observable_from_counts}
\end{equation}
where \(O(b)\) is the eigenvalue of the observable corresponding to the measured configuration \(b\).

\subsubsection{Density Evolution}

The particle density at lattice site \(j\) is
\begin{equation}
\hat n_j = \hat n_{j\uparrow}+\hat n_{j\downarrow}
% \hat n_{2j}+\hat n_{2j+1}.
\end{equation}
Using Eq.~\eqref{eq:number_pauli},
\begin{equation}
\hat n_j
=1-\frac{1}{2}
\left(
\hat Z_{2j}+\hat Z_{2j+1}
\right).
\end{equation}
The density obtained from the hardware counts is therefore
\begin{equation}
\langle \hat n_j\rangle
=\sum_b p_b
\left(
b_{2j}+b_{2j+1}
\right),
\label{eq:density_counts}
\end{equation}
or, equivalently,
\begin{equation}
\langle \hat n_j\rangle
=1-\frac{1}{2}
\left(
\langle Z_{2j}\rangle+
\langle Z_{2j+1}\rangle
\right).
\end{equation}
Repeating this procedure at each evolution time provides the site-resolved density evolution.

\subsubsection{Imbalance}

To quantify the memory of the initial charge-density-wave configuration, we use the charge imbalance
\begin{equation}
\mathcal I(t)
=\frac{
N_{\mathrm{even}}(t)-N_{\mathrm{odd}}(t)
}{
N_{\mathrm{even}}(t)+N_{\mathrm{odd}}(t)
},
\label{eq:imbalance_def}
\end{equation}
where
\begin{align}
N_{\mathrm{even}}
&=
\sum_{j,\mathrm{even}}
\left(
n_{j\uparrow}+n_{j\downarrow}
\right),
\\
N_{\mathrm{odd}}
&=
\sum_{j\,\mathrm{odd}}
\left(
n_{j\uparrow}+n_{j\downarrow}
\right).
\end{align}
For a fixed-particle-number sector with total particle number \(N\), Eq.~\eqref{eq:imbalance_def} becomes
\begin{equation}
\mathcal I(t)
=
\frac{1}{N}
\sum_{j=0}^{L-1}
(-1)^j
\langle \hat n_j(t)\rangle.
\label{eq:imbalance_density}
\end{equation}

For each measured bitstring \(b\), the corresponding imbalance is
\begin{equation}
\mathcal I(b)
=
\frac{1}{N}
\sum_{j=0}^{L-1}
(-1)^j
\left(
b_{2j}+b_{2j+1}
\right),
\end{equation}
and the hardware estimate is
\begin{equation}
\boxed{
\mathcal I(t)
=
\sum_b p_b\,\mathcal I(b).
}
\label{eq:imbalance_counts}
\end{equation}

In terms of Pauli operators,
\begin{equation}
\hat{\mathcal I}
=
\frac{1}{N}
\sum_{j=0}^{L-1}
(-1)^j
\left[
1-\frac{
\hat Z_{2j}+\hat Z_{2j+1}
}{2}
\right].
\end{equation}
For an even number of lattice sites, the identity contributions cancel, yielding
\begin{equation}
\hat{\mathcal I}
=
-\frac{1}{2N}
\sum_{j=0}^{L-1}
(-1)^j
\left(
\hat Z_{2j}+\hat Z_{2j+1}
\right).
\label{eq:imbalance_pauli}
\end{equation}
Thus, the imbalance can be obtained entirely from single-qubit \(Z\)-expectation values measured in the computational basis.

For the \(L=6\), 12-qubit system considered in the main text, with \(N=6\),
\begin{align}
\mathcal I(t)
=
\frac{1}{12}
\big[
&
\langle Z_2\rangle+\langle Z_3\rangle
+\langle Z_6\rangle+\langle Z_7\rangle
+\langle Z_{10}\rangle+\langle Z_{11}\rangle
\nonumber\\
&
-\langle Z_0\rangle-\langle Z_1\rangle
-\langle Z_4\rangle-\langle Z_5\rangle
-\langle Z_8\rangle-\langle Z_9\rangle
\big],
\label{eq:imbalance_12q}
\end{align}
where the convention is chosen such that the initially occupied sublattice has \(\mathcal I(0)=1\).

\subsubsection{Second-order participation entropy}

The participation entropy characterizes the spreading of the many-body state over the computational, or equivalently Fock, basis. Since the quantum processor directly samples the computational-basis probability distribution, this quantity can be obtained without evaluating additional Pauli operators.

The participation entropy of order \(q\) is defined as
\begin{equation}
S_q^{\mathrm{PE}}
=
\frac{1}{1-q}
\ln
\left(
\sum_b p_b^q
\right).
\end{equation}
In this work, we consider the second-order participation entropy,
\begin{equation}
\boxed{
S_2^{\mathrm{PE}}(t)
=
-\ln
\left[
\sum_b p_b(t)^2
\right].
}
\label{eq:participation_entropy}
\end{equation}
% The quantity
% \begin{equation}
% \mathrm{IPR}(t)
% =
% \sum_b p_b(t)^2
% \end{equation}
% is the inverse participation ratio, and consequently
% \begin{equation}
% S_2^{\mathrm{PE}}=-\ln(\mathrm{IPR}).
% \end{equation}

For a state localized entirely in a single computational-basis configuration, \(p_b=1\) for one bitstring and zero otherwise, giving \(S_2^{\mathrm{PE}}=0\). If the probability is uniformly distributed over \(M\) configurations, \(p_b=1/M\), then \(S_2^{\mathrm{PE}}=\ln M\). Thus, an increase in \(S_2^{\mathrm{PE}}\) indicates spreading of the wave function over a larger number of many-body configurations.
Unlike the density and imbalance, \(S_2^{\mathrm{PE}}\) is not the expectation value of a linear quantum operator. It is a nonlinear functional of the measured probability distribution and is therefore evaluated directly from the complete set of measured probabilities according to Eq.~\eqref{eq:participation_entropy}.

\subsubsection{Quantum Fisher information}

We characterize many-body correlations through the quantum Fisher information associated with the staggered charge operator
\begin{equation}
\hat O_c
=
\sum_{j=0}^{L-1}
(-1)^j
\left(
\hat n_{j\uparrow}+\hat n_{j\downarrow}
\right).
\label{eq:qfi_generator}
\end{equation}
For a pure state, the quantum Fisher information corresponding to the generator \(\hat O_c\) is
\begin{equation}
\boxed{
F_Q(t)
=
4
\left[
\langle \hat O_c^2(t)\rangle
-
\langle \hat O_c(t)\rangle^2
\right].
}
\label{eq:qfi}
\end{equation}

For every measured bitstring \(b\), the staggered charge is evaluated as
\begin{equation}
O_c(b)
=
\sum_{j=0}^{L-1}
(-1)^j
\left(
b_{2j}+b_{2j+1}
\right).
\label{eq:staggered_charge_bitstring}
\end{equation}
The quantities are then obtained directly from the hardware probability distribution:
\begin{align}
\langle \hat O_c\rangle
&=
\sum_b p_b O_c(b),
\label{eq:qfi_first_moment}
\\
\langle \hat O_c^2\rangle
&=
\sum_b p_b [O_c(b)]^2.
\label{eq:qfi_second_moment}
\end{align}
Substitution into Eq.~\eqref{eq:qfi} gives
\begin{equation}
\boxed{
F_Q(t)
=
4
\left[
\sum_b p_b O_c(b)^2
-
\left(
\sum_b p_b O_c(b)
\right)^2
\right].
}
\label{eq:qfi_counts}
\end{equation}

% It is important that \(O_c(b)\) is squared for each individual measurement outcome before performing the statistical average. In general,
% \begin{equation}
% \langle O_c^2\rangle
% \neq
% \langle O_c\rangle^2,
% \end{equation}
% and their difference represents the fluctuations of the staggered charge.

Using Eq.~\eqref{eq:number_pauli}, the staggered charge operator can also be expressed in terms of Pauli-\(Z\) operators. For even \(L\),
\begin{equation}
\hat O_c
=
-\frac{1}{2}
\sum_{j=0}^{L-1}
(-1)^j
\left(
\hat Z_{2j}+\hat Z_{2j+1}
\right).
\end{equation}
Consequently, \(\langle O_c^2\rangle\) contains both single-qubit \(Z\) expectation values and two-qubit correlations of the form
\begin{equation}
\langle Z_p Z_q\rangle.
\end{equation}
All such quantities are diagonal in the computational basis and can therefore be extracted from the same measurement dataset without requiring additional measurement bases.

% For a fixed total particle number \(N\), the staggered charge and imbalance are related through
% \begin{equation}
% O_c=N\mathcal I,
% \end{equation}
% such that
% \begin{equation}
% F_Q
% =
% 4N^2
% \left(
% \langle \mathcal I^2\rangle
% -
% \langle \mathcal I\rangle^2
% \right).
% \label{eq:qfi_imbalance_relation}
% \end{equation}
% Thus, while the imbalance quantifies the average memory of the initially imposed density modulation, the quantum Fisher information probes the fluctuations and correlations associated with the same staggered charge distribution.

% In summary, the computational-basis measurement outcomes obtained from the quantum processor provide all the information required to evaluate the observables considered here. The local density and imbalance are obtained as linear expectation values of occupation operators, the participation entropy is calculated from the full probability distribution, and the quantum Fisher information is obtained from the first and second moments of the staggered charge distribution.

\section{Comparison between quantum-hardware and exact-diagonalization results}
\label{sec:hardware_ed_comparison}

To obtain the accuracy of the quantum-hardware calculations, the quantum hardware results are compared with corresponding results obtained using the exact diagonalization of the t-FHM. For each value of the Stark-field strength \(\Delta_{0}\) and evolution time \(t\), the hardware value reported below corresponds to the mean of three independent executions of the quantum circuit on the quantum processor. The same observables were independently evaluated from the state obtained by exact diagonalization. To quantify the deviation of the hardware results from the \textit{exact-diagonalization} (ED) reference, we define the \textit{percentage fractional difference} (PFD) as
\begin{equation}
\mathrm{PFD}
=
100 \times
\frac{
O_{\mathrm{HW}}-O_{\mathrm{ED}}
}{
O_{\mathrm{ED}}
},
\label{eq:pfd}
\end{equation}
where \(O_{\mathrm{HW}}\) and \(O_{\mathrm{ED}}\) denote the observable obtained from the quantum hardware and exact diagonalization, respectively. The PFD is not defined when the exact reference value is zero. 

Table~\ref{tab:hw_ed_all_observables} compares all the observables obtained from the quantum processor with the exact-diagonalization results. For Imbalance in the strongest Stark field, \(\Delta_{0}=15\), the hardware results closely follow the exact dynamics throughout the considered time interval, with deviations below approximately \(2.5\%\). Larger relative deviations appear for the weaker fields at times where the imbalance approaches or crosses zero, for which the relative-error measure becomes particularly sensitive to small absolute differences.
 For QFI at \(t=0\), since the ED value is zero, the percentage fractional difference cannot be assigned at this point. The small nonzero values obtained from hardware at \(t=0\) arise from noise present in the device.
For \(\Delta_{0}=1\) and \(3\), the hardware values reproduce the exact-diagonalization results reasonably closely over the full evolution interval. For the strongly localized regime, \(\Delta_{0}=15\), larger relative deviations are observed because the QFI itself remains considerably smaller than in the weak-field regime, making the relative deviation more sensitive to relatively small changes. Similar calculations are also done for the second order participation entropy.
Overall, the quantum-hardware calculations reproduce the qualitative and, in most cases, quantitative behavior obtained from exact diagonalization. 

For the DW initial state, the half-chain particle number \(N_{\rm half}\) obtained from the quantum hardware shows very good agreement with the exact-diagonalization results over the full evolution interval, Table \ref{tab:dw_nhalf}. The percentage fractional difference remains below \(0.5\%\) for all values of \(\Delta_0\) and \(t\), with the largest deviation occurring at \(\Delta_0=3\) and \(t=1.25\). In the strongly tilted regime, \(\Delta_0=15\), the agreement is particularly close, with deviations below approximately \(0.05\%\), indicating that the hardware accurately reproduces the suppression of particle transport across the two halves of the system.

\begin{table*}[htbp]
\centering
\scriptsize
\setlength{\tabcolsep}{3.2pt}
\renewcommand{\arraystretch}{1.08}
\caption{Comparison of the observables obtained from quantum hardware (HW) and exact diagonalization (ED) for the charge density wave state. The hardware values are averaged over three independent executions.
A dash indicates that the percentage fractional difference (PFD)
is undefined because the corresponding ED value is zero.}
\label{tab:hw_ed_all_observables}

\begin{tabular}{cc
rrr
rrr
rrr}
\hline
& &
\multicolumn{3}{c}{Imbalance ,$\mathcal{I}$} &
\multicolumn{3}{c}{QFI, $F_Q$} &
\multicolumn{3}{c}{Participation entropy, $S_2^{\mathrm{PE}}$}
\\
\cline{3-5}
\cline{6-8}
\cline{9-11}

$\Delta_{0}$ & $t$
& HW & ED & PFD (\%)
& HW & ED & PFD (\%)
& HW & ED & PFD (\%)
\\
\hline

1 & 0.00
& 0.9994 & 1.0000 & -0.06
& 0.0309 & 0.0000 & --
& 0.0039 & 0.0000 & --
\\

1 & 0.25
& 0.7813 & 0.8043 & -2.86
& 8.2757 & 7.9897 & 3.58
& 1.3335 & 1.2254 & 8.82
\\

1 & 0.50
& 0.3222 & 0.3481 & -7.44
& 16.5048 & 16.5685 & -0.38
& 4.2224 & 4.1115 & 2.70
\\

1 & 0.75
& -0.0786 & -0.0922 & -14.75
& 14.3241 & 13.9810 & 2.45
& 5.0945 & 5.0459 & 0.96
\\

1 & 1.00
& -0.2821 & -0.3078 & -8.36
& 11.4331 & 11.0170 & 3.78
& 4.5636 & 4.5020 & 1.37
\\

1 & 1.25
& -0.2799 & -0.2660 & 5.21
& 14.1119 & 13.7306 & 2.78
& 4.7175 & 4.7039 & 0.29
\\
\hline

3 & 0.00
& 0.9995 & 1.0000 & -0.05
& 0.0232 & 0.0000 & --
& 0.0029 & 0.0000 & --
\\

3 & 0.25
& 0.7956 & 0.8107 & -1.86
& 8.0467 & 7.7739 & 3.51
& 1.2076 & 1.1838 & -2.01
\\

3 & 0.50
& 0.3995 & 0.4199 & -4.86
& 16.6222 & 16.0574 & 3.52
& 3.6465 & 3.7072 & -1.64
\\

3 & 0.75
& 0.0985 & 0.1031 & -4.42
& 15.2082 & 15.4956 & -1.85
& 4.9937 & 4.9344 & 1.20
\\

3 & 1.00
& -0.0356 & -0.0307 & 16.04
& 12.3472 & 12.2577 & 0.73
& 4.8733 & 4.8082 & 1.35
\\

3 & 1.25
& 0.0216 & 0.0170 & -27.05
& 11.003 & 11.2754 & 2.43
& 4.7113 & 4.6139 & -2.77
\\
\hline

15 & 0.00
& 0.9994 & 1.0000 & -0.06
& 0.0307 & 0.0000 & --
& 0.0038 & 0.0000 & --
\\

15 & 0.25
& 0.9348 & 0.9404 & -0.60
& 3.0963 & 2.7299 & 13.42
& 0.3973 & 0.3623 & 9.66
\\

15 & 0.50
& 0.9658 & 0.9815 & -1.60
& 1.1758 & 0.8801 & 33.60
& 0.1125 & 0.1114 & -0.98
\\

15 & 0.75
& 0.9442 & 0.9676 & -2.42
& 1.8250 & 1.5255 & 19.64
& 0.2460 & 0.2112 & -8.64
\\

15 & 1.00
& 0.9561 & 0.9579 & -0.18
& 2.2162 & 1.9742 & 12.26
& 0.2589 & 0.2540 & -1.92
\\

15 & 1.25
& 0.9606 & 0.9795 & -1.93
& 1.1734 & 1.0033 & 16.95
& 0.1213 & 0.1216 & 0.24
\\

\hline
\end{tabular}
\end{table*}

\begin{table}[htbp]
\centering
\caption{Comparison of the half-chain particle number $N_{\rm half}$ 
for the domain-wall initial state obtained from quantum hardware (HW)
and exact diagonalization (ED). The hardware values are averaged over 
three independent executions.}
\label{tab:dw_nhalf}
\begin{tabular}{ccccc}
\hline
$\Delta_{0}$ & $t$ & HW & ED & PFD (\%) \\
\hline
1  & 0.00 & 5.9975 & 6.0000 & -0.0417 \\
1  & 0.25 & 5.8681 & 5.8789 & -0.1831 \\
1  & 0.50 & 5.5447 & 5.5595 & -0.2659 \\
1  & 0.75 & 5.1499 & 5.1538 & -0.0753 \\
1  & 1.00 & 4.7729 & 4.7801 & -0.1506 \\
1  & 1.25 & 4.4929 & 4.4945 & -0.0366 \\
\hline
3  & 0.00 & 5.9977 & 6.0000 & -0.0383 \\
3  & 0.25 & 5.8783 & 5.8813 & -0.0518 \\
3  & 0.50 & 5.5895 & 5.5936 & -0.0729 \\
3  & 0.75 & 5.2761 & 5.2867 & -0.2005 \\
3  & 1.00 & 5.0793 & 5.0831 & -0.0756 \\
3  & 1.25 & 5.0126 & 5.0333 & -0.4109 \\
\hline
15 & 0.00 & 6.0000 & 6.0000 &  0.0000 \\
15 & 0.25 & 5.9583 & 5.9610 & -0.0459 \\
15 & 0.50 & 5.9946 & 5.9947 & -0.0025 \\
15 & 0.75 & 5.9697 & 5.9707 & -0.0159 \\
15 & 1.00 & 5.9807 & 5.9819 & -0.0203 \\
15 & 1.25 & 5.9841 & 5.9849 & -0.0136 \\
\hline
\end{tabular}
\end{table}

% \section{List of changes:}
% \begin{enumerate}
%     \item the delta thing
%     \item give the PFD values as a table (I mean the table of average values and then the exact diagonalization)
%     % \item derivation of the tilted Fermi-Hubbard model and attach it in the supplemental material.
%     % \item Maybe we can add the results without post-processing 
% \end{enumerate}

\end{document}